\documentclass[structabstract]{aa}  
\usepackage{graphicx}

\newcommand{\HII}{H{\sc ii}}
\newcommand{\Ak}{$A_{K_{\rm{s}}}$}

\newcommand{\Ks}{$K_{\rm{s}}$}
\newcommand{\micron}{$\mu$m}
\newcommand{\msun}{M$_{\sun}$}

\usepackage{natbib}
\usepackage{txfonts}
\begin{document}
   \title{Deep near-infrared imaging of W3 Main: constraints on stellar cluster formation.\thanks{Based on data acquired using the Large Binocular Telescope (LBT). The LBT is an international collaboration among institutions in Germany, Italy and the United States. LBT Corporation partners
are: LBT Beteiligungsgesellschaft, Germany, representing the Max Planck Society, the Astrophysical Institute Potsdam, and Heidelberg University; Istituto Nazionale di AstroÞsica, Italy; The University of
Arizona on behalf of the Arizona university system; The Ohio State
University, and The Research Corporation, on behalf of The University
of Notre Dame, University of Minnesota and University of Virginia.}}

   \author{A. Bik
          \inst{1}
          \and
	A. Stolte
           \inst{2}
           \and
           M. Gennaro
           \inst{1,3}           
           \and
	      W. Brandner
		\inst{1}
		\and
           D. Gouliermis
           \inst{1,4}
           \and
          B. Hu\ss mann
           \inst{2}
           \and
	  E. Tognelli
	  \inst{5}
	  	\and
		B. Rochau
		\inst{1}
		\and
		Th. Henning
		\inst{1}
		\and
		A. Adamo
		\inst{1}
		\and
		H. Beuther
		\inst{1}	
		\and
		A. Pasquali
		\inst{6}
		\and
		Y. Wang
		\inst{7}		
          }
   \institute{Max-Planck-Institut f\"ur Astronomie, K\"onigstuhl 17, 69117 Heidelberg, Germany\\
              \email{abik@mpia.de}
      \and
  	Argelander Institut f\"ur Astronomie, Auf dem H\"ugel 71, 53121 Bonn, Germany
	\and 
	Space Telescope Science Institute, 3700 San Martin Dr., Baltimore, MD 21218, USA
	\and	
	Universit\"at Heidelberg, Zentrum f\"ur Astronomie, Institut f\"ur Theoretische Astrophysik, Albert-Ueberle-Str. 2, 69120 Heidelberg, Germany
	\and
	Department of Physics, University of Pisa, Largo Bruno Pontecorvo 3, 56127, Pisa, Italy
		\and
		Universit\"at Heidelberg, Zentrum f\"ur Astronomie, Astronomisches Rechen Institut,  M\"onchhofstrasse 12 - 14, 69120 Heidelberg, Germany
	\and
	Purple Mountain Observatory, Chinese Academy of Sciences, 210008, Nanjing, PR China
}

   \date{Received; accepted }

 
  \abstract
   {Embedded clusters like W3 Main are complex and dynamically evolving systems that represent an important phase of the  star formation process.}
   {We aim at the characterization of the entire stellar content of W3 Main in a statistical sense to identify possible differences in evolutionary phase of the stellar populations  and find clues about the formation mechanism of this massive embedded cluster.}
   {Deep $JH$\Ks\ imaging is used to derive the disk fraction, \Ks-band luminosity functions and mass functions for several subregions in W3 Main. A two dimensional completeness analysis using artificial star experiments is applied as a crucial ingredient to assess realistic completeness limits for our photometry.}
   {We find an overall disk fraction of   7.7 $\pm$2.3\%, radially varying from 9.4$\pm$ 3.0 \% in the central 1 pc to 5.6 $\pm$ 2.2 \% in the outer parts of W3 Main. The mass functions derived for three subregions are consistent with a Kroupa and Chabrier mass function. The mass function of IRSN3 is complete down to 0.14 \msun\ and shows a break at M $\sim$ 0.5 \msun. }
   {We interpret the higher disk fraction in the center as evidence for a younger age of the cluster center. We find that the evolutionary sequence observed in the low-mass stellar population is consistent with the observed age spread among the massive stars.  An analysis of the mass function variations does not show evidence for mass segregation. W3 Main is currently still actively forming stars, showing that the ionizing feedback of OB stars is confined to small areas ($\sim$0.5 pc). The FUV feedback might be influencing large regions of the cluster as suggested by the low overall disk fraction.}


   \maketitle

\section{Introduction}\label{sec:intro}

Embedded clusters play an important role in the star formation process: it is in these embedded clusters where most, or even all, young stars are born. In such star-forming regions,  stars interact  via stellar feedback and dynamical interactions with each other and with the surrounding interstellar medium (ISM). Therefore understanding the physical processes operating on small (tenths of pc) and large scales (several tens of pc) becomes fundamental to perceive the impact of star formation on global galactic scales.

Near-infrared as well as mid-infrared observations have been very effective in uncovering the stellar content of many embedded clusters in our Galaxy \citep[e.g.][]{Lada03,Gutermuth09}. These observations uncovered a suite of different morphologies of embedded clusters. They range from very compact and centrally concentrated clusters to loose associations. \citet{Elmegreen08,Elmegreen10} proposed that star formation is hierarchically structured, and explain the variety of observed morphologies by suggesting that associations and starburst clusters follow the self-similar distribution of dense gas in their parent cloud, driven by self gravity and turbulence.

To derive constraints on the formation mechanisms of  embedded clusters, a detailed study of the stellar and gas content provides vital information. The identification of the massive stars  as well as the molecular and ionized gas gives insights into feedback mechanisms acting in embedded clusters. The characterization of the lower-mass stars reveals important information on the evolutionary phase of the stellar population. The characterization of the stellar population in a statistical sense using disk fractions, luminosity functions, and mass functions allows us to place constraints on the cluster formation and its early evolution.

The fraction of objects possessing a circumstellar disk is a strong function of the age of the stellar population.  Most of the work on disk fractions is carried out using the $L$-band or the Spitzer/IRAC 3.6 \micron\ band as the excess indicator \citep[e.g.][]{HaischDisk01,Hernandez08,Alcala08}. These studies showed a strong decline of the disk fraction with  cluster age. The observed fraction changes from almost 100\% for the youngest clusters ($<$ 1 Myr) to 50\% after a few Myr and almost zero after $\sim$ 5 -7 Myr, thus constraining the inner-disk dissipation time \citep{Hernandez08}. For disk fractions derived from $JH$\Ks\ data alone, the disk fractions are typically a factor 0.6  of those derived from $L$-band data \citep{Yasui09,Yasui10}.

A  tool to study the entire stellar population of a stellar cluster to very high extinction and low stellar masses, deeper than possible with multi-band photometry, is the $K$-band luminosity function (KLF). \citet{Muench00} demonstrated that the shape of the KLF is influenced by the underlying initial mass function (IMF), cluster age and star formation history as well as the mass-to-luminosity relation of the pre-main-sequence (PMS) models.

The IMF is an important statistical tool to study the stellar content and properties of embedded star clusters. Studies of many different stellar systems have revealed that the IMF is very uniform over a large range of different environments \citep[e.g.][]{Bastian10}. However, variations in the present day mass function (MF) have been observed. The dynamical evolution of stellar clusters can result in the removal of the lowest mass stars and the migration of the most massive ones to the center \citep[e.g.][]{Allison09,Harfst10}. This mass segregation will lead to a flattening of the observed stellar mass function in the cluster center and a steepening in the outskirts.

The embedded cluster W3 Main is particularly interesting in this respect.  W3 Main is part of the large star formation complex W3/W4/W5 located in the Perseus arm in the outer Galaxy \citep{Megeath08}. The complex is located at a distance of 1.95 $\pm$ 0.04 kpc \citep{XuW306} and  consists of a series of OB associations still partly embedded inside their natal molecular cloud.  The W3 region is considered  the youngest of the regions, still associated with a Giant Molecular Cloud. W3 consists of two optically visible \HII\ regions IC 1795 and NGC 896 with an age of 3-5 Myr \citep{OeyW305, Roccatagliata11}. W3 Main is one of the regions located inside the molecular cloud wrapped around these \HII\ regions.  

High-resolution radio continuum observations of W3 Main revealed that a large variety of different HII regions are present, ranging from diffuse  to  hyper-compact \HII\ regions \citep{WynnWilliams71,Harris76,Claussen94,Tieftrunk97}. Their observed sizes range from few pc  to 0.01 pc in diameter. This can be interpreted as an evolutionary sequence, where the hyper-compact \HII\ regions represent the youngest regions and the diffuse \HII\ regions the more evolved objects \citep{Wood89,Kurtz02}. Hyper-compact and ultra-compact \HII\ regions have estimated ages of less than 100,000 yr \citep{Hoare07} and are ionized by the most recently formed OB stars. 

The candidate massive stars inside the \HII\ regions were identified using near-infrared imaging  of W3 Main \citep{Ojha04}. By means of $K$-band spectroscopy, their massive star nature was confirmed and the spectral types of these massive stars have been determined \citep{Bik12} such that the age of the most massive star (IRS2) could be derived. Based on this age determination (2-3 Myr) and the presence of the very young hyper-compact HII regions, \citet{Bik12} concluded that W3 Main has been actively forming stars over the last 2-3 Myr. 

Spectroscopy is a very powerful way to classify both the high- and low-mass stellar content of stellar clusters \citep[e.g.][]{Ostarspec05,Bik10,Hanson10}.  It is, however, limited to the brightest objects and typically, even with the most advanced techniques, only several tens of objects can be characterized. To extend the characterization of the entire stellar content and to detect possible different  evolutionary phases in the low-mass stellar populations, statistical methods based on near-infrared photometry are more suited. 
Previous imaging studies of W3 Main  mainly focussed on the central parts around IRS5. A dense, very deeply embedded, stellar cluster has been identified by \citet{Megeath96,Megeath05} and studied in detail.  
Deep adaptive optics observations have revealed brown dwarfs in the outskirts of W3 Main \citep{Ojha09}, while a large population of PMS stars was detected using sensitive X-ray observations \citep{Feigelson08}.

In this paper we present our detailed photometric analysis of the stellar content of W3 Main by analyzing deep  $JH$\Ks\ imaging 
 of the embedded cluster. The paper is organized as follows: in Section \ref{sec:observations} we discuss the data reduction and photometry. Section \ref{sec:stellarpop} presents the detailed analysis of the stellar content of W3 Main where we derive disk fractions and construct \Ks-band luminosity functions and mass functions. In Section \ref{sec:discussion} we discuss the results by comparing it to previous results  and discuss their implication for possible cluster formation scenarios  and summarize the conclusions in Section \ref{sec:conclusions}. 

   \begin{figure}
   \centering
  \includegraphics[width=\columnwidth]{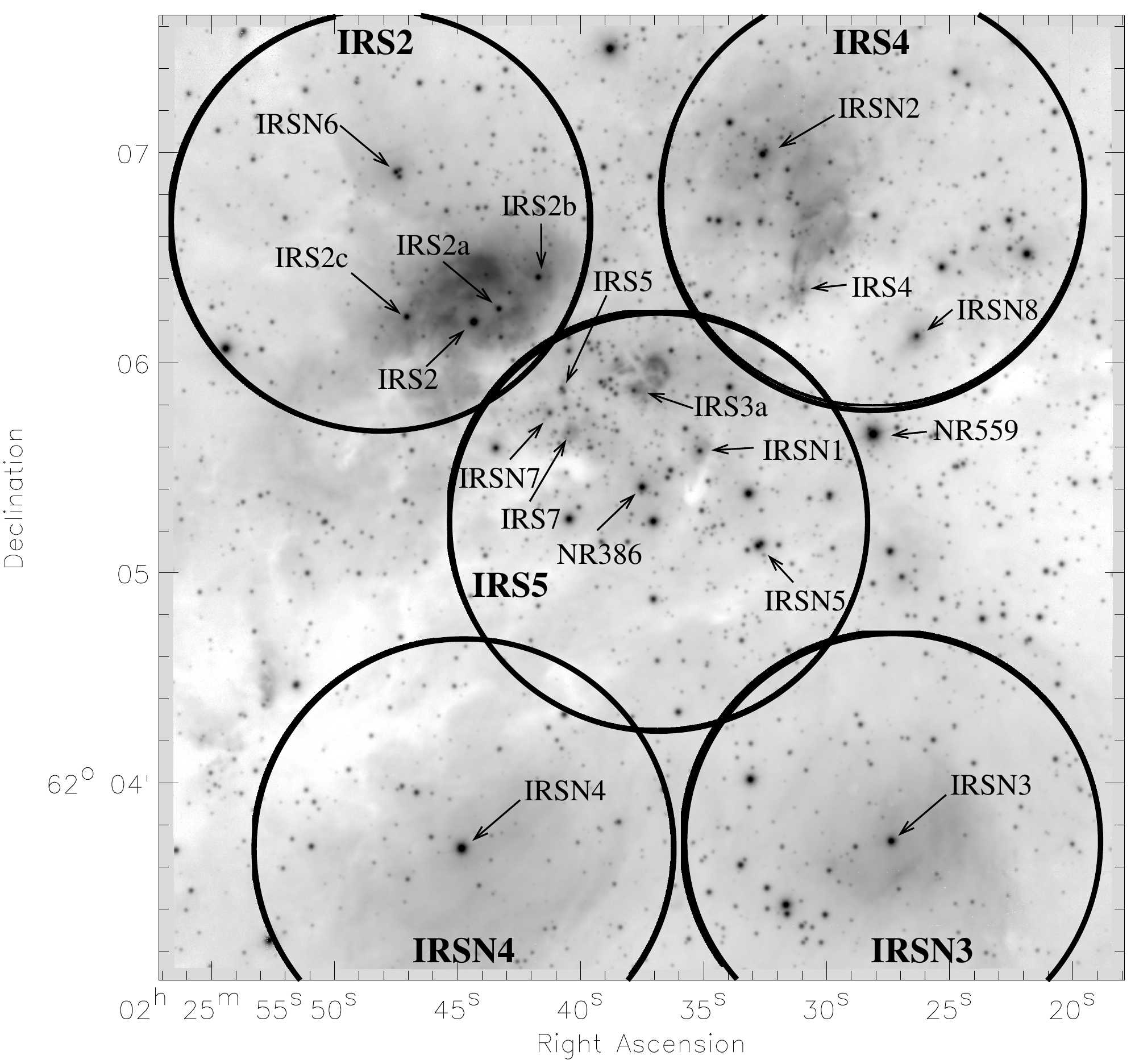}
   \caption{Overview \Ks\ map of the W3 Main region. Location of the regions discussed in the tex and selected for plotting the CMDs (Fig. \ref{fig:cmd_HII}). The areas are circular, covering the bright \HII\ regions with a radius of 1\arcmin.  The regions centered on IRS N3 and N4 cover the diffuse \HII\ regions W3 J and W3 K respectively. The region on IRS2  contains  the compact \HII\ region W3A, and the area  around IRS4 encompasses the \HII\ regions W3 D, W3 C and W3 H. The 1\arcmin circle  around IRS5 covers the youngest \HII\ regions: W3 B, W3 E , W3 F and W3 M. \label{fig:regions}}
       \end{figure}

\section{Observations and data reduction}\label{sec:observations}

\subsection{Photometry}\label{sec:photometry}

   \begin{figure}
   \centering
  \includegraphics[width=\columnwidth]{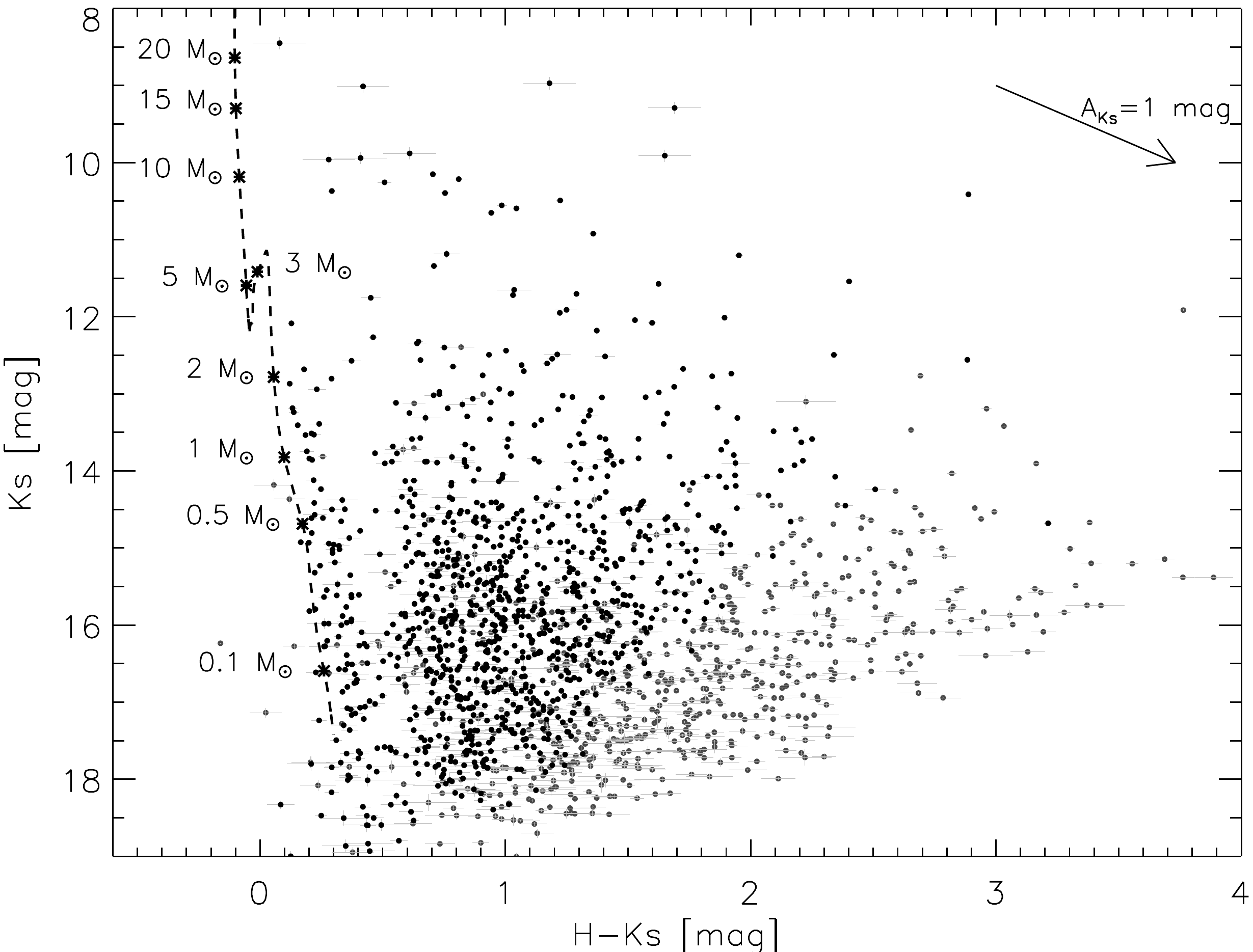}
   \caption{$H$-\Ks\ vs \Ks\ color-magnitude diagram of W3 Main. The grey data points represent the sources detected in $H$ and \Ks, while the black data points are the sources detected in $JH$\Ks. The very red sources are not detected anymore in the $J$-band due to their high-extinction which affects the detection in the $J$-band the strongest.  Only sources with magnitude errors less than 0.1 mag are plotted. Overplotted is the extinction vector for \Ak = 1 mag from the extinction law of  \citet{Nishiyama09}. The dashed line shows the combined main sequence and PMS isochrone of 2 Myr for a distance of 2 kpc. Above 5 M$_{\sun}$, the main sequence models of \citet{Brott11} with vrot = 0 are used, while for lower masses the models have been calculated by \citet{Tognelli11}. }\label{fig:cmd}
       \end{figure}

Near-infrared $JH$\Ks\ imaging of the stellar content of W3 Main was obtained with  LUCI1 \citep[LBT NIR spectroscopic Utility with Camera and Integral-Field Unit for Extragalactic Research,][]{Ageorges10,Seifert10,Buschkamp10} mounted at the Gregorian focus of the Large Binocular Telescope (LBT) on Mount Graham, Arizona \citep{Hill06}. The  imaging observations of W3 Main were performed on December 17 - 19, 2009 and are described in detail in \citet{Bik12}. The data  were obtained with the N3.75 camera, providing a 0.12 \arcsec pixel scale with a  field of view of 4\arcmin $\times$ 4\arcmin.

The $JH$\Ks\ images of W3 Main are centered on $\alpha$(2000) = 02$^h$25$^m$37.5$^s$,  $\delta$(2000) = +62$^{\circ}$05\arcmin 24\arcsec. Additionally, we observed a control field that was used for both  sky subtraction and  as  photometric control field. This field was centered on  $\alpha$(2000) = 02$^h$25$^m$41.0$^s$,  $\delta$(2000) = +62$^{\circ}$12\arcmin 32\arcsec.  The measured image quality of the images in the $J$- and $H$-band was 0.6\arcsec\, while  an image quality of 0.9\arcsec\  of the \Ks\ image was derived.

The imaging data were reduced using standard IRAF\footnote{IRAF is distributed by the National Optical Astronomy Observatory, which is operated by the Association of Universities for Research in Astronomy, Inc., under cooperative agreement with the National Science Foundation.} routines.  Apart from the standard data reduction steps, including dark and flat field correction as well as  sky subtraction, we corrected the frames for geometric distortion before combining them to the final image.  Fig. \ref{fig:regions} shows the reduced \Ks\ image with the massive stars identified in \citet{Bik12} annotated.
Point spread function (PSF) fitting photometry on the $JH$\Ks\ images was performed using \emph{starfinder} \citep{Diolaiti00}. The PSF was constructed from 22 bright and isolated stars in the image and the final PSF was created iteratively after selecting and subtracting faint neighboring sources from the image.   Realistic photometric errors are determined by splitting the imaging dataset in two subsets and repeating the photometry procedure described above.   The catalogues of the different  bands are matched allowing a maximum offset of 4 pixels (0.48\arcsec). 

The absolute photometric calibration is performed using the 2MASS photometry \citep{Skrutskie06}.  We selected only stars with magnitudes of 10 mag $<$ $J$ $<$ 16.5 mag, 10 mag $<$ $H$ $<$ 14.5 mag, and 9.5 mag $<$ \Ks $<$ 14.0 mag to avoid saturated stars and to  ensure good 2MASS photometry for the fainter stars. We selected only the 2MASS sources with quality flag "AAA". We visually inspected the matches between 2MASS and LUCI stars to remove  mismatches due to different spatial resolution and sensitivities. Additionally, we only selected those stars with $|$ mag(LUCI) - mag(2MASS) $|$ $<$ 0.5 mag from the median zero point of all stars in each filter. This led to a selection of 54 stars in J, 49 stars in H, and 53 stars in Ks as final calibrators. The resulting uncertainties in the zero points are  $\sigma_{J_{\rm{zpt}}}$ =  0.022 mag, $\sigma_{H_{\rm{zpt}}}$ = 0.014 mag, and $\sigma_{K_{\rm{s,zpt}}}$=  0.018 mag. No color-terms in the zero points were found. Additional details on the photometry can be found in  \citet{Bik12}.
 This procedure results in 2659 sources detected in \Ks\, 2331 sources in the $H$-band and 1506 in $J$. The $H$ and \Ks\ matched catalogue consists of 1968 sources and the $JH$\Ks\ catalogue contains 1282 sources. Note that the $J$-band image is slightly offset from the $H$ and \Ks\ pointings, meaning that the $JH$\Ks\ catalogue covers a smaller area than the $H$\Ks\ catalogue, resulting in less objects. The resulting $H$-\Ks\ vs. \Ks\ color-magnitude diagram is plotted in Fig. \ref{fig:cmd}.

\subsection{Completeness}\label{sec:completeness}

   \begin{figure*}
   \centering
  \includegraphics[width=18cm]{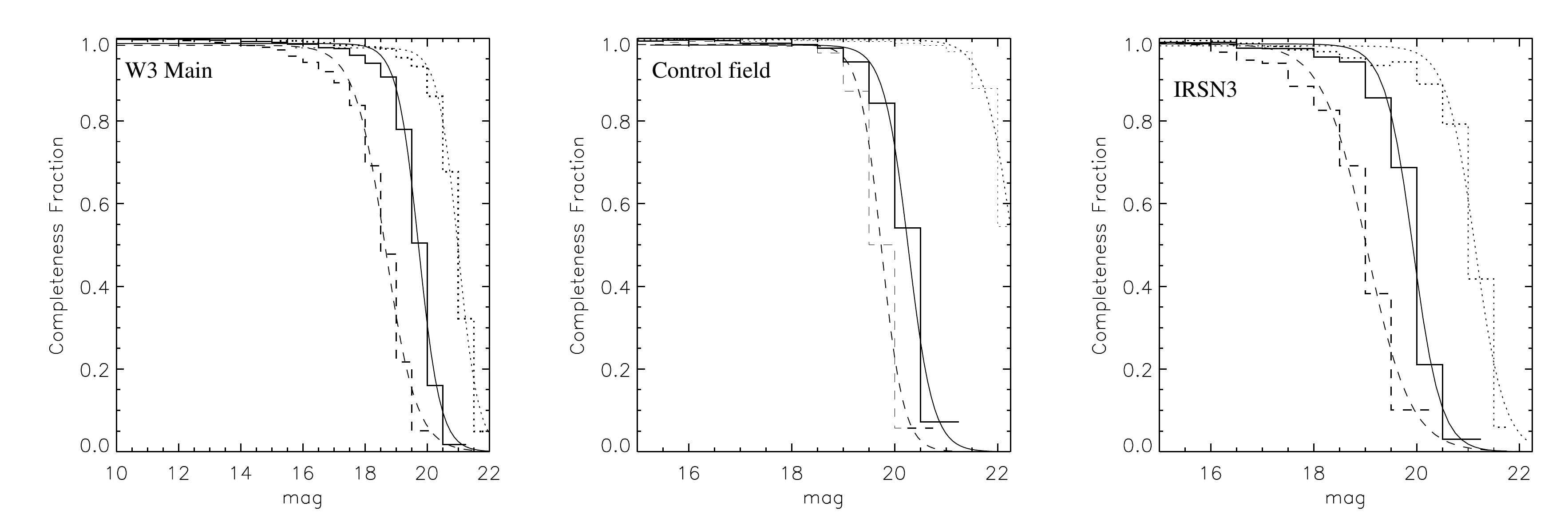}
   \caption{Fraction of the photometric completeness as a function of the observed magnitude in W3 Main (left), the control field (middle) and the IRSN3 subregion of W3 Main (right). The dashed lines show the average completeness fraction for the \Ks-band, the solid lines the $H$-band and the dotted lines shows the $J$-band completeness fraction. The resulting average 50 \% completeness limits are given in Table \ref{tab:regions}.} \label{fig:completeness_plots}
       \end{figure*}

The completeness of our photometry is determined by several factors. 
 Apart from the sensitivity of the observations, the photometric completeness in embedded clusters is strongly affected by  crowding and bright background due to extended emission associated with the \HII\ regions. Additionally, variable extinction causes the loss of the reddest and most obscured stars. Due to these effects the completeness will vary strongly over the observed field as especially the strong emission of the \HII\ regions is very localized.  By performing artificial star experiments we can characterize the photometric incompleteness due to the sensitivity, crowding and strong background emission. The effect of the variable
extinction can not be determined and is discussed in  Sect. \ref{sec:extinction}.

We adopted the \emph{starfinder} implementation \citep{Hussmann12} of the procedure described in \citet{Gennaro11} to carry out the artificial star experiments and create  2D completeness maps. Artificial stars were randomly added each 0.5 magnitude bin in such an amount that the crowding properties were not changed. The stars were added using the PSF obtained from the real data.  PSF fitting photometry with \emph{starfinder} was performed to recover the artificially added stars.  We transformed the positions of the artificial stars in the $H$-band to the $J$ and the \Ks\ bands to calculate the completeness in those two bands.  

We performed two experiments, covering the observed magnitude range between $H$ = 9.5 and 21 mag. The magnitude ranges for \Ks\ and $J$ were chosen such that $H$-\Ks\ = 1.0 mag and $J$-$H$ = 1.5 mag, characteristic colors of the PMS population of the less reddened subregions in W3 Main (Fig. \ref{fig:cmd}).  For the bright stars ($H$ = 9.5 - 14 mag, \Ks = 8.5 - 13 mag, $J$ = 11 - 15.5 mag) we added  in every 0.5 magnitude bin randomly 100 stars  (7\%, 4\% and 4\% of the total detected sources in respectively $J$, $H$, \Ks) and repeated this 20 times to get a total of 2000 stars without changing the crowding properties of the image.  To obtain a higher spatial resolution sampling for the fainter stars ($J$ = 15.5 - 23 mag,  $H$ = 14 - 20.5 mag and \Ks\ = 13 - 21.5 mag), we performed a second set of artificial star experiments using 8000 stars per magnitude bin (i.e. 250 stars per run, 32 runs).
 
We combine the result of these two sets to construct a two dimensional map of the completeness for every magnitude bin by interpolation from the list of artificial star positions.   By combining the two runs, the completeness fraction is calculated between $J$ = 11 - 23 mag, $H$ = 9.5 - 21 mag and \Ks\ = 8.5 - 21.5 mag. An average value for each pixel was created by averaging over the 16 nearest neighbors using a Kriging interpolation \citep[Apendix A2,][]{Gennaro11}. With an effective detector area ($A_{\mbox{eff}}$) of $6.08\times 10^6$ pixels$^2$ and $N_{\mbox{bin}}$ = 2000 stars per magnitude bin we obtain an effective spatial resolution for the bright stars \citep[using Appendix A1 from][]{Gennaro11} of:

\begin{equation}
d_{\mbox{eff}} = \sqrt{\frac{A_{\mbox{eff}}}{\pi N_{\mbox{bin}}}} \times \sqrt{\nu} \approx 124 \mbox{ pixel} = 15 \arcsec,
\end{equation}

where $\nu$ is the number of nearest neighbors used in the interpolation.    For the brightest magnitude bins between $H$ = 9.5 and 14 mag, the spatial resolution of the completeness analysis is less critical as the completeness is still close to unity. For the fainter stars, adding 8000 stars results in an effective resolution of 7.4\arcsec ($\sim 10$ times the FWHM). The resulting average completeness curves are shown in Fig \ref{fig:completeness_plots}. The completeness curves show an average 50\% completeness level at $J$ = 20.7 mag, $H$ = 19.5 mag and \Ks\ = 18.4 mag.

   \begin{figure}
   \centering
  \includegraphics[width=\columnwidth]{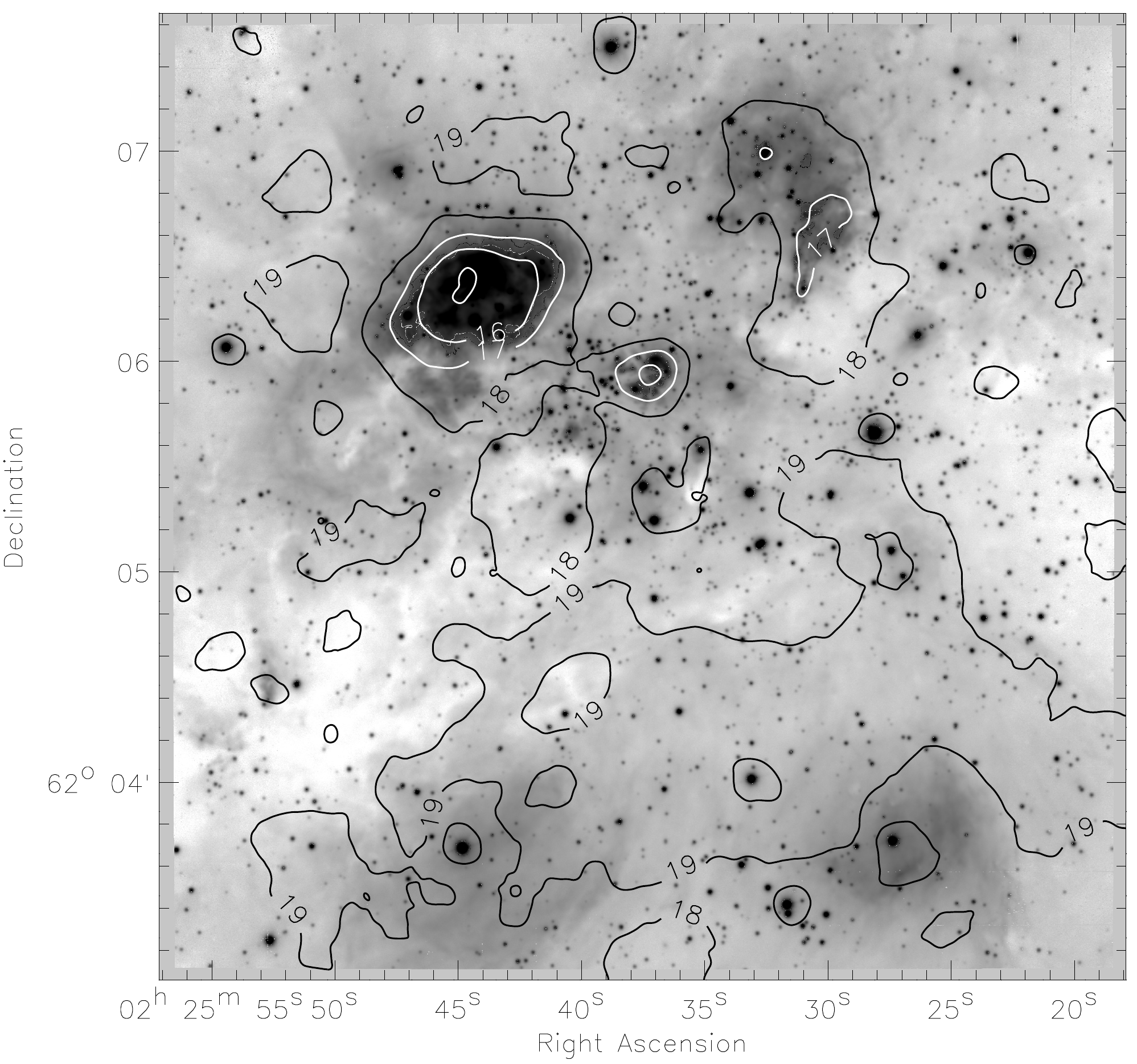}
   \caption{Map of the 50 \% completeness limits in \Ks. Plotted in grayscale is the \Ks-band image in logarithmic stretch. Overplotted as contours is the 50 \% completeness map as derived from the artificial star experiments. The values are indicated with the contour levels. The white contour levels mark the lowest completeness magnitudes: \Ks = 15, 16, 17 mag. The 50 \% completeness ranges from \Ks = 19 mag in the southern part of the region to \Ks = 15 mag towards W3A, demonstrating a dramatic change in photometric completeness. } \label{fig:completeness_map}
    \end{figure}

For each pixel, we fitted the completeness value obtained per magnitude bin as a function of magnitude using a sigmoid function \citep{Gennaro11}. 
In this way we were able to obtain the 50\% completeness limit in each band and for each pixel.  The results  for the \Ks\ band are shown in Fig. \ref{fig:completeness_map} as contours overplotted on the \Ks\ image. The derived 50\% completeness magnitudes range from \Ks\ = 14.9 mag in the brightest parts of W3 A to \Ks\  = 19.7 mag in the least crowded areas.  This demonstrates the need of a 2D completeness analysis for studying the stellar content of such a complex region. In the $H$-band, the 50\% completeness varies between $H$ = 17.2 and 20.4 mag, while the $J$-band completeness ranges from $J$ = 18.5 to 21.6 mag. 

\subsection{Control field}\label{sec:control}

\begin{figure}
   \centering
  \includegraphics[width=\columnwidth]{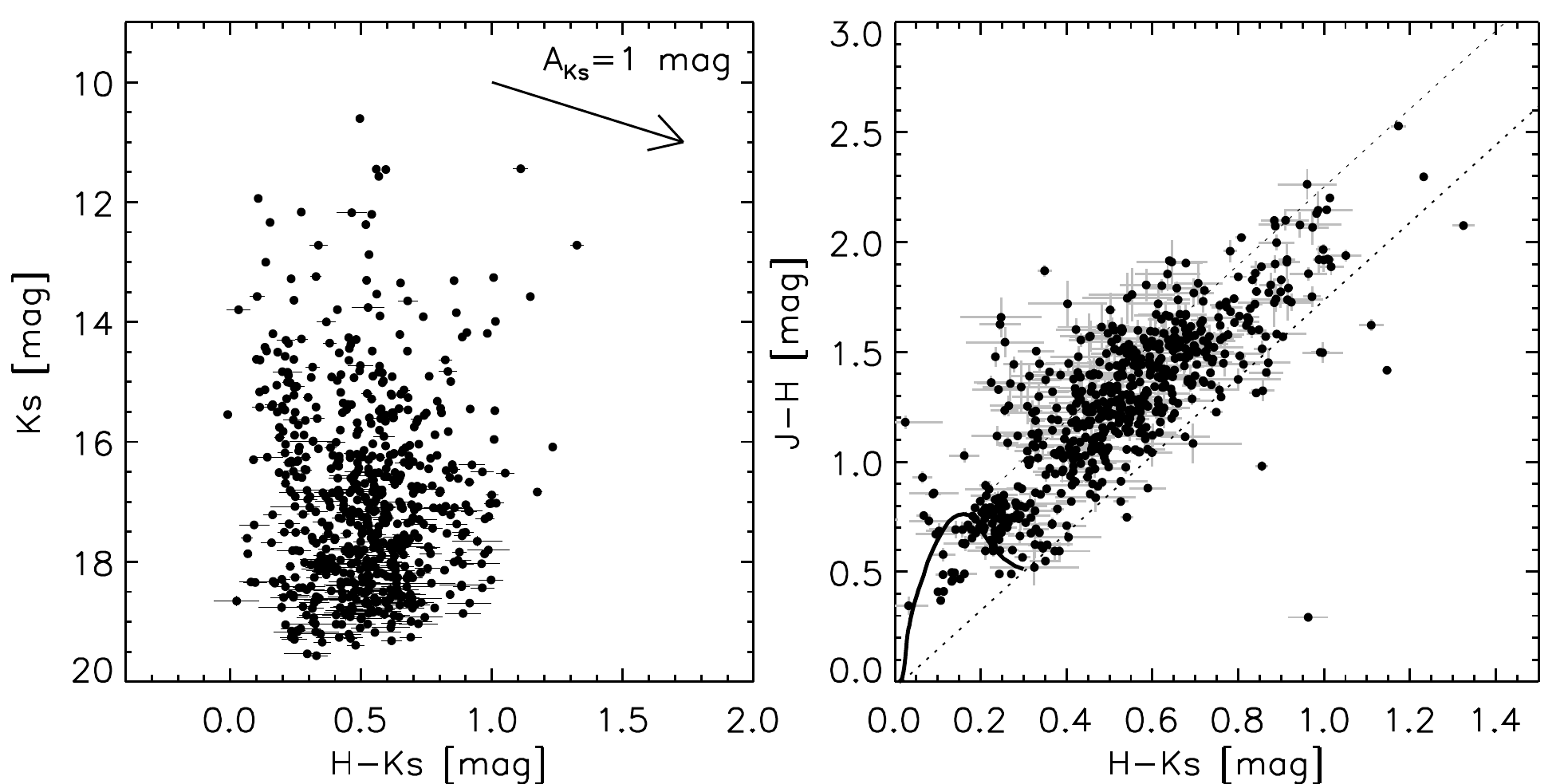}
   \caption{\emph{Left}: $H$-\Ks\ vs. \Ks\ color-magnitude diagram of the control field offset from W3 Main, centered on  $\alpha$(2000) = 02$^h$25$^m$41.0$^s$,  $\delta$(2000) = +62$^{\circ}$12\arcmin 32\arcsec. \emph{Right}: $H$-\Ks\ vs $J$-$H$ color-color diagram of the same field. The solid line shows a 2 Myr isochrone and the diagonal dotted lines outline the area where reddened main sequence stars are located.  \label{fig:cmd_control}}
       \end{figure}

To obtain an idea of the  contamination by field stars located in front or behind W3 Main, we used the field observed for sky correction as the control field.  We performed photometry identical to the \emph{starfinder} photometry of the science images. The absolute photometry was obtained after calibration with the 2MASS catalogue and the individual bands were matched with a 4 pixel tolerance. Fig. \ref{fig:cmd_control} shows the resulting $H$- \Ks  vs. \Ks\ color-magnitude diagram (CMD) and $J$-$H$ vs. $H$-\Ks\ color-color diagram (CCD).   

We computed the photometric incompleteness in the same way as for the science field by adding artificial stars in the images. We limited the completeness analysis to calculating only the average completeness, using 2000 stars per 0.5 magnitude bins, as  no extended emission is present and the 50 \% completeness limit is assumed to be constant over the image. This analysis results in a 50 \% completeness magnitude of $J$ = 22.0 mag, $H$ = 20.0 mag and \Ks\ = 19.5 mag.

\begin{table*}
\caption{Summary of selected subregions  in W3 Main and their photometric completeness limits.}             
\label{tab:regions}      
\centering                   
\begin{tabular}{r c c r r r c c r c}       
\hline\hline                
Name\tablefootmark{a} & $\alpha$ (J2000) & $\delta$ (J2000) & sources & Area &  50 \% $J$ & 50\% $H$&  50\% \Ks&\HII\ regions \\ 
&	(h m s) & ($^\circ$\ \arcmin\  \arcsec) & &  \arcmin$^2$ & mag & mag & mag   &	 \\
\hline                        
W3 Main & 02:25:37.5	& +62:05:24     & 1681   &  19.9/18.9\tablefootmark{b}  &18.6    &17.2   & 14.9  & All \\
IRS5		&02:25:36.9	& +62:05:15	& 358     & 3.14                                      &19.5    & 18.7  & 15.8  & M, F, E\\
IRS4		&02:25:28.1	& +62:06:47     & 252     & 2.98/2.64\tablefootmark{b}   &19.7    & 19.0  & 16.8  & Ca, C, D, H	\\	
IRS2		&02:25:48.2	& +62:06:40     & 325     &  3.10/2.82\tablefootmark{b}  &18.6    &17.2   & 14.9  & A\\
IRS N3	&02:25:27.4	& +62:03:43     & 205     & 2.70                                      & 19.5   & 19.0  &17.5   & J\\
IRS N4	&02:25:44.8	& +62:03:41	& 184     & 2.65                                      & 19.7   &19.0   & 17.2  & K \\ 
Control	&02:25:41.0     & +62:12:32     & 854     &17.37                                     & 22.0   & 20.0  & 19.5  & --- \\ 
\hline                                  
\end{tabular}
\tablefoot{
\tablefoottext{a} The locations of the selected regions and the dominant infrared sources (IRS) in respect to the observed field-of-view are shown in the map of  Fig. \ref{fig:regions}.\\
\tablefoottext{b} The $J$-band pointing was slightly offset to the South from the $H$- and \Ks-band pointing. This results in the subregions IRS4 and IRS2 being slightly less well covered in $JH$\Ks\ than they are in $H$\Ks.
}
\end{table*}

\section{Results}\label{sec:stellarpop}

W3 Main harbors several \HII\ regions in different evolutionary stages, ranging from the youngest hyper-compact \HII\ regions to the more evolved diffuse \HII\ regions \citep{Tieftrunk97}. It consists of several generations of star formation with a difference in age of a few Myr \citep{Bik12}.  In this section, we use the photometry of the stellar content to derive statistical properties to constrain the evolutionary phase of the stellar population as a function of position in W3 Main. We derive the extinction map and study the clustering properties of W3 Main.   We also separate the data in
different subregions and derive disk fraction, \Ks-band luminosity functions and mass functions as a function of location in the cluster.

The derivation of the cluster's extinction and stellar masses have been performed by means of theoretical isochrones. We use  models  from the Pisa PMS database\footnote{The Pisa PMS database is available at the url: http://astro.df.unipi.it/stellar-models/} \citep{Tognelli11} for masses in the range 0.1 -  7.0 M$_\odot$, while for higher masses we adopted the models computed by Brott et al. (2011) for galactic abundances and zero rotation velocity. The two sets of isochrones overlap in the mass region (4 - 7.8 \msun), hence we decided to joint them close to the main-sequence at a mass of approximatively 5 M$_\odot$. Theoretical models have been transformed to the observational plane by adopting the GAIA spectra \citep{Brott05} for $T_\mathrm{eff} \le 10\,000$ K, and the \citet{Castelli03} spectra for $T_\mathrm{eff} > 10\,000$ K.  As an example the 2 Myr isochrone is overlayed in Fig. \ref{fig:cmd}. This isochrone was chosen as reference as it matches the age of the most massive star in W3 Main \citep[IRS2,][]{Bik12}.

\subsection{Extinction}\label{sec:extinction}

The  CMD of W3 Main shows all the sources detected in at least 2 bands with a photometry error less than 0.1 mag (Fig. \ref{fig:cmd}). The sample of stars detected only in  $H$\Ks\ (gray dots) covers a color range of up to $H$-\Ks\ = 3 - 4 mag, while the sample detected in $JH$\Ks\ (black dots) is less complete towards the high extinction part and only the brightest objects are detected in $J$.  This shows that  extinction has a strong effect on the detection rate,  especially in the $J$-band. For the massive stars the extinctions were derived in \citet{Bik12} and vary between  \Ak\ = 0.3 mag and \Ak\ = 5.9 mag.

\begin{figure}
   \centering
  \includegraphics[width=\columnwidth]{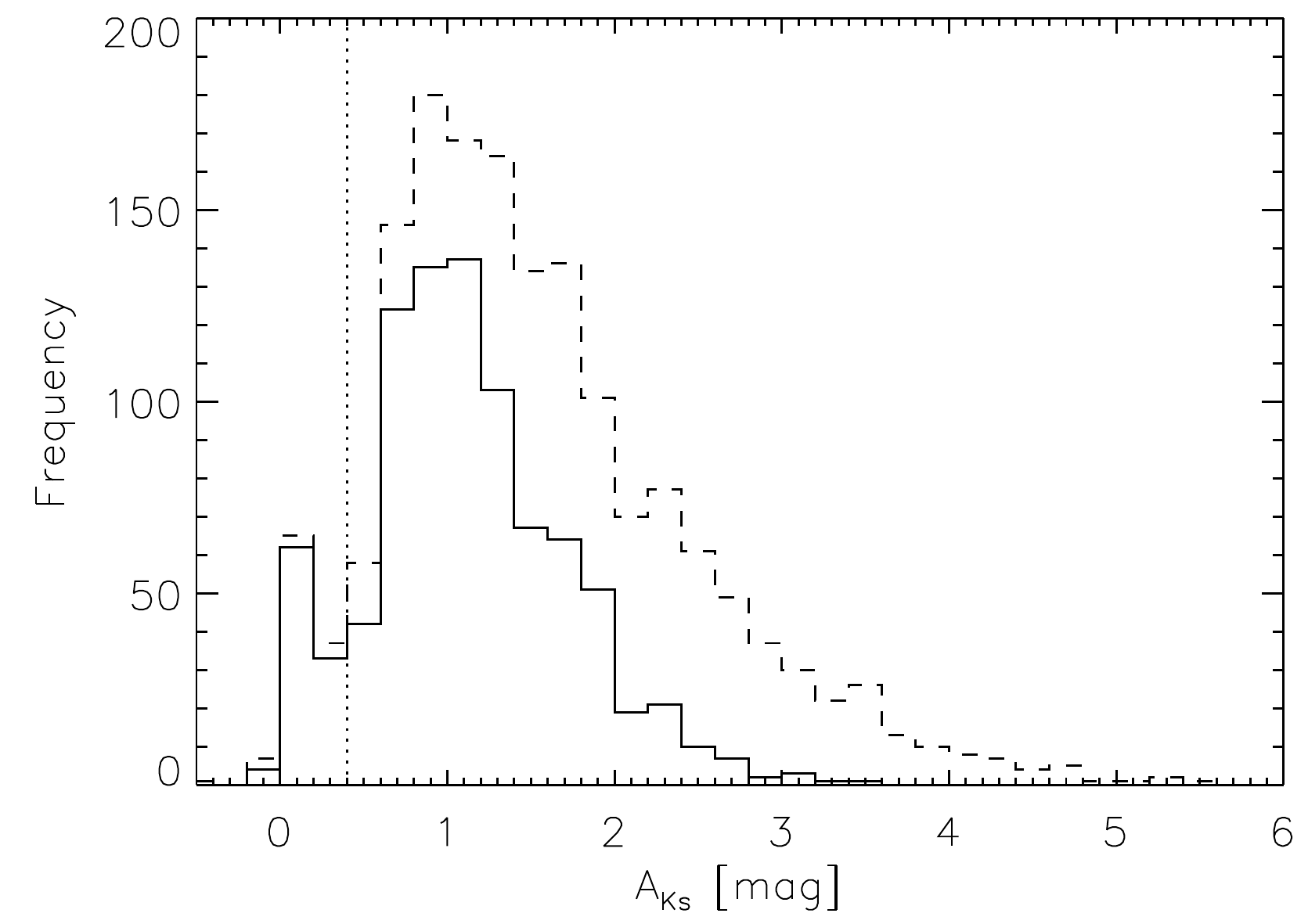}
   \caption{Histogram of the extinction towards the stars detected in W3 Main with a bin size of 0.2 mag. Using the \citet{Nishiyama09}  extinction law, the stars are de-reddened in the CMD on to a 2 Myr main sequence and PMS isochrone. Changing the isochrone to 1 or 3 Myr does not change significantly the shape of the histogram.
The solid line shows the histogram of the extinctions derived towards stars with $JH$\Ks\ detections, while the dashed line represents the extinction histogram of all the  $H$\Ks\ sources (including the $JH$\Ks\ sources). The vertical dotted line, denotes \Ak = 0.4 mag and shows  the division line between the foreground and cluster population.} \label{fig:hist}
       \end{figure}

To derive the extinction of the entire stellar population, we de-redden the sources in the $H$, $H$-\Ks\ CMD (Fig. \ref{fig:cmd}) to their main-sequence or PMS location assuming an age of 2 Myr. 
We removed the infrared-excess sources from the $JH$\Ks\ sample as their color is not only caused by extinction but also by circumstellar emission (see Sect. \ref{sec:excess} for more details).  For the sources detected only in $H$\Ks, we could not remove the contamination of the infrared-excess sources as only one color is available. Using the extinction law of \citet{Nishiyama09} we derive the extinction towards all the point sources.  Fig. \ref{fig:hist} shows the histogram of the extinction distributions. The solid line represents the extinction distribution of the $JH$\Ks\ detections, while the dashed histogram shows the distribution of the sources with  $H$\Ks\ detections (including the $JH$\Ks\ sources). It becomes evident that the $H$\Ks\ sample probes much higher extinctions than the $JH$\Ks\ sample due to the $J$-band detection limit.

The extinction histogram shows a peak at \Ak\ = 0.1 mag,  caused by foreground stars at a distance of less than 2 kpc. \citep{XuW306}. These stars can also be seen as the blue sequence in the CMD with $H$-\Ks\ $< $ 0.5 mag (Figs. \ref{fig:cmd},  \ref{fig:cmd_control}).
This value is consistent with the mean interstellar galactic reddening towards this part of the Galaxy. \citet{Joshi05} derived values for $E(B-V)$ $\sim$ 0.35 mag per kpc towards the direction of W3 Main. This translates to \Ak\ = 0.12/kpc mag using the \citet{Rieke85} extinction law. To translate the optical extinction to \Ak, we use the \citet{Rieke85} extinction law, as no optical values are given for the \citet{Nishiyama09} extinction law.

 At \Ak\ $\sim$ 0.4 mag, the histogram rises strongly and peaks at \Ak\ = 1 mag. This suggests that the total foreground extinction towards W3 Main would be around \Ak\ = 0.4 mag. The remaining extinction is caused by local reddening inside the cluster. The shape of the high-extinction end of the histogram is affected by our incomplete detection of the reddest stellar content. Note that at low \Ak\ ($<$ 1.0 mag) more sources are present in the $H$\Ks\ sample than in the $JH$\Ks\ sample, this is due to the slightly shifted $J$-band pointing.

To study the spatial distribution of the extinction we created an extinction map. We removed the foreground sources with \Ak$<$ 0.4 mag and created a 2D extinction map of the cluster area from the individual extinction values towards each star. To have the largest amount of sources and probe the highest extinctions we used the $H$\Ks\ sample. This catalogue contains 1509 sources in the LUCI field of view. After removing 76 sources with \Ak\ $<$ 0.4 mag, we are left with 1433 sources to construct the extinction map. To create the extinction map we first calculated, for every star position in the catalogue, the average local extinction using its ten nearest neighbors. After that the data points were interpolated on a regular grid with the size of the \Ks\ image using a linear interpolation. 

   \begin{figure}
   \centering
  \includegraphics[width=\columnwidth]{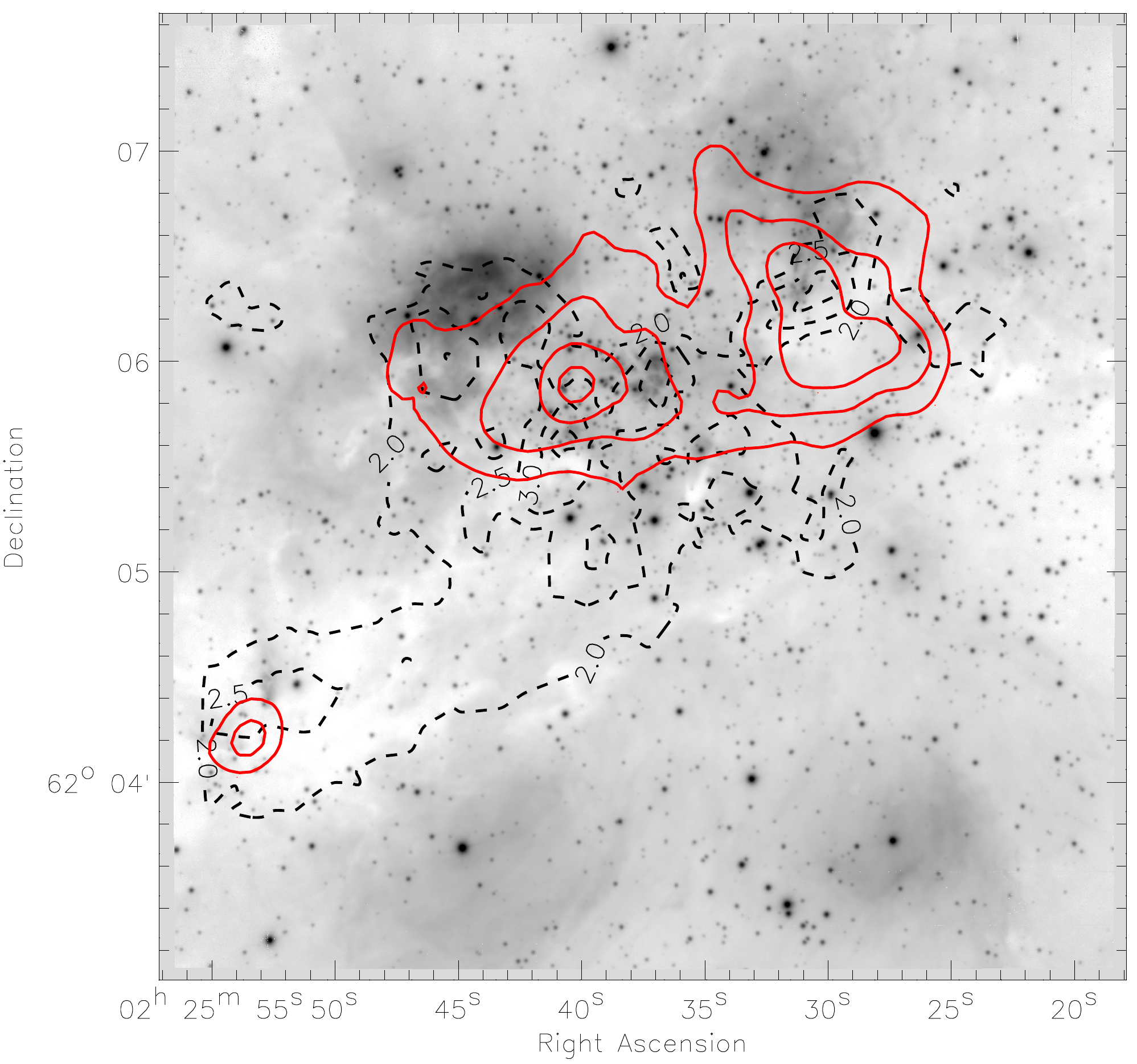}
   \caption{Extinction map towards W3 Main. The image shows the negative \Ks-band image with logarithmic stretch. Overplotted in dashed black contours is the extinction map derived from the $H$-\Ks\ colors of the stars using the extinction law of \citet{Nishiyama09}. Contours at \Ak\ = 2, 2.5, 3.0 and 3.5 mag are shown. The red contours display  450 \micron\  continuum emission observed by SCUBA \citep{Ladd93,DiFrancesco08},  highlighting the location of the dense and cold dust. The contours of 10\%, 20\%, 40\% and 80\% of the maximum flux value (6.4 Jy/beam) are shown. The dust emission peaks around IRS5 and IRS4.}\label{fig:akmap}
       \end{figure}


Fig. \ref{fig:akmap} shows the resulting extinction map. The \Ak\ = 2 - 3.5 mag contours are overplotted in the \Ks\ image. The extinction map shows three areas with an average extinction higher than \Ak = 3 mag, two associated with the cluster around IRS5 in the center of the image, and one area associated with IRS4, in the north-west of the image. Towards the two diffuse \HII\ regions, W3 J (around IRSN3) and W3 K (around IRSN4), the derived extinction is very low, consistent with what is found for the ionizing stars of these regions \citep{Bik12}.  Comparison between the extinction map and sub-mm data shows that the highest extinction areas coincide with two dense molecular cores (W3 East towards IRS5 and W3 west towards IRS4) detected at 450 \micron\ with SCUBA \citep{Ladd93,Moore07} as well as Herschel \citep{RiveraIngraham13}. A third peak, south of IRS4 is detected in the 450 \micron\ SCUBA map (SWS3) and is harboring several starless cores \citep{Wang12}. In the central $\sim$ 30\arcsec\ of these molecular cores, the extinction becomes so high that the stars become so reddened and fall below our sensitivity limits, remaining undetected  even in the \Ks-band. This explains the lower \Ak\ values measured towards the center of the IRS5 molecular core.

\subsection{Sub clustering}

   \begin{figure}
   \centering
  \includegraphics[width=\columnwidth]{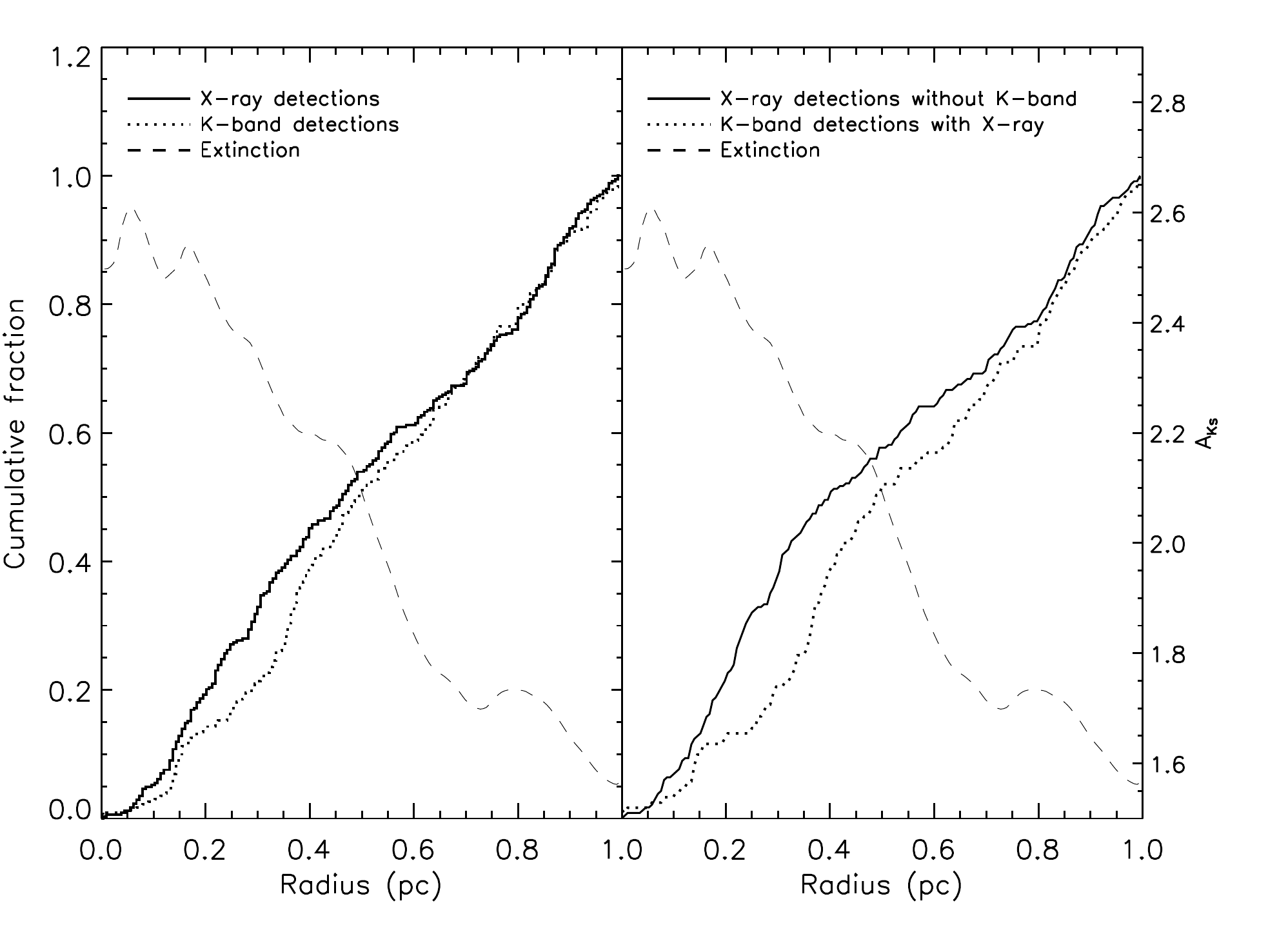}
   \caption{Cumulative radial distribution of the sources in the central parsec of W3 Main. The location of the center is chosen to be the position of IRS5. \emph{Left:}  The solid line shows the cumulative distribution of the X-ray sources of \citet{Feigelson08} and the dotted line shows  the distribution of the \Ks-only catalogue. \emph{Right:} The solid line shows the distribution of the X-ray sources without \Ks\ counterpart and the dotted line shows the cumulative distribution of the \Ks-band sources matched with the X-ray catalogue. The dashed line in both figures is the radial extinction profile derived from radially averaging the extinction map derived in Fig. \ref{fig:akmap}.} \label{fig:cumulative}
       \end{figure}

The near-infrared stellar distribution in W3 Main shows large spatial variations, which could be interpreted as sub-clustering inside the region. However, in the previous subsection, we observed strong spatial variations in the near-IR derived extinction, with two heavily extinguished areas towards IRS5 and IRS4. In order to see whether these substructures in the stellar distribution are real or caused by extinction toward the line of sight, we compare the radial distribution of near-infrared stars with respect to the distribution of X-ray sources contained in the catalogue by \citep{Feigelson08}. 

X-ray observations allow the detection of the low-mass PMS star population  \citep{Feigelson99}. Moreover, extinctions of up to \Ak\ = 15 - 50 mag can be probed using sensitive X-ray observations  \citep{Getman05}. Therefore, in young, still partially embedded, stellar clusters, the distribution recovered from X-ray detections is closer to the real stellar distribution than the one observed in the near-infrared. In the case of W3 Main, the X-ray observations show a stellar distribution nearly spherically centered on IRS5 with a slight density enhancement towards IRS4  \citep{Feigelson08}. To derive a clear map of the morphology of W3 main we compare cumulative radial distributions of X-ray and near-IR sources inside the central parsec of W3 Main, centered on IRS5. A larger radius would fall beyond the borders of our near-infrared images.

We have matched the positions of the X-ray sources with our \Ks-only catalogue using the \emph{match\_xy.pro} routine of the TARA\footnote{http://www2.astro.psu.edu/xray/docs/TARA/} software package \citep{Broos10}.  This routine takes into account the positional accuracies of the stars, which can vary strongly for the X-ray sources, ranging from $\sim$ 0.02\arcsec\ to 2.4\arcsec\ for some sources. The positional errors on the infrared detections are a lot smaller and negligible compared to those of the X-ray positions.  A total of 351 X-ray sources are detected inside the 1 pc radius around IRS5  as well as 225 \Ks\ band sources above \Ks\ = 14.9 mag.  The limit of \Ks = 14.9 mag was chosen to ensure that the \Ks\ sample is complete across the central parsec so that the spatial variations are not caused by locally varying photometric incompleteness (see Sect. \ref{sec:completeness}, Fig \ref{fig:completeness_map}). After the matching, 52\% of the \Ks\ sources possess an X-ray counterpart and 64\% of the X-ray sources  have no  \Ks\ counterpart brighter than 14.9 mag.

Fig. \ref{fig:cumulative} shows the cumulative radial distributions of the central 1 pc area around IRS5.  In the left panel, the radial distribution of  all X-ray sources (solid line) is compared to that of  all \Ks\ sources with \Ks\ brighter than the \Ks\ = 14.9 mag completeness limit (dotted line). The cumulative distribution of the \Ks\ sources is corrected for photometric completeness using the 2D completeness calculations. 
Also plotted is the radial extinction profile derived from the extinction map created in Sect. 3.1 (dashed line). We radially averaged the extinction map with the center of the map on IRS5. Additionally, in the right panel of Fig. \ref{fig:cumulative} we show the cumulative distributions of the \Ks\ sources with an X-ray counterpart (dotted) and the X-ray sources without a detectable \Ks\ counterpart (solid). 

These graphs show that the radial distribution of  X-ray sources is more concentrated towards the center than the radial distribution of \Ks-band sources. This becomes more clear when comparing the radial profiles of the X-ray sources without \Ks\ counterpart with that of  the \Ks\ sources with an X-ray counter part. The two curves deviate in the inner 0.4 pc where the extinction is highest. The X-ray profile shows a higher concentration of sources in the center while the radial profile of the \Ks\ sources shows a much shallower profile.  Only at the outskirts of the cluster (r $>$ 0.7 pc), where the extinction becomes lower, the curves have the same slope, tracing the same population.   When applying a Kolmogorov-Smirnov (KS) test to compare the curves of the \Ks\ sample with X-ray counterpart and the X-ray without \Ks\ counterpart, a 0.06 probability is found that they are drawn from the same distribution. This supports the suggestion that the extinction profile changes the stellar distribution observed in the \Ks-band. Therefore we conclude that the sub-clustering visible in the near-infrared is dominated by extinction variations and that the real stellar distribution is better observed in the X-rays, as much higher extinctions can be probed with X-rays then with near-infrared imaging.

To estimate whether the fore- and background contamination can have a dominant effect on the above comparison we use the control field to estimate the level of contamination. We apply the method described in Sect. \ref{sec:excess}. 
After applying the same selection as for the science sample (\Ks $<$ 14.9 mag), 47 contaminating sources with \Ks $<$ 14.9 are expected across the same area as covered by the W3 Main science field.  By scaling that to the central 1 pc area, we expect 24 fore- or background sources in our sample. This is only 10 percent of the total sample used, therefore we do not expect this to influence the derived sub clustering properties.

\subsection{Stellar populations in W3 Main}\label{sec:HIIregions}

   \begin{figure*}
   \centering
  \includegraphics[width=18cm]{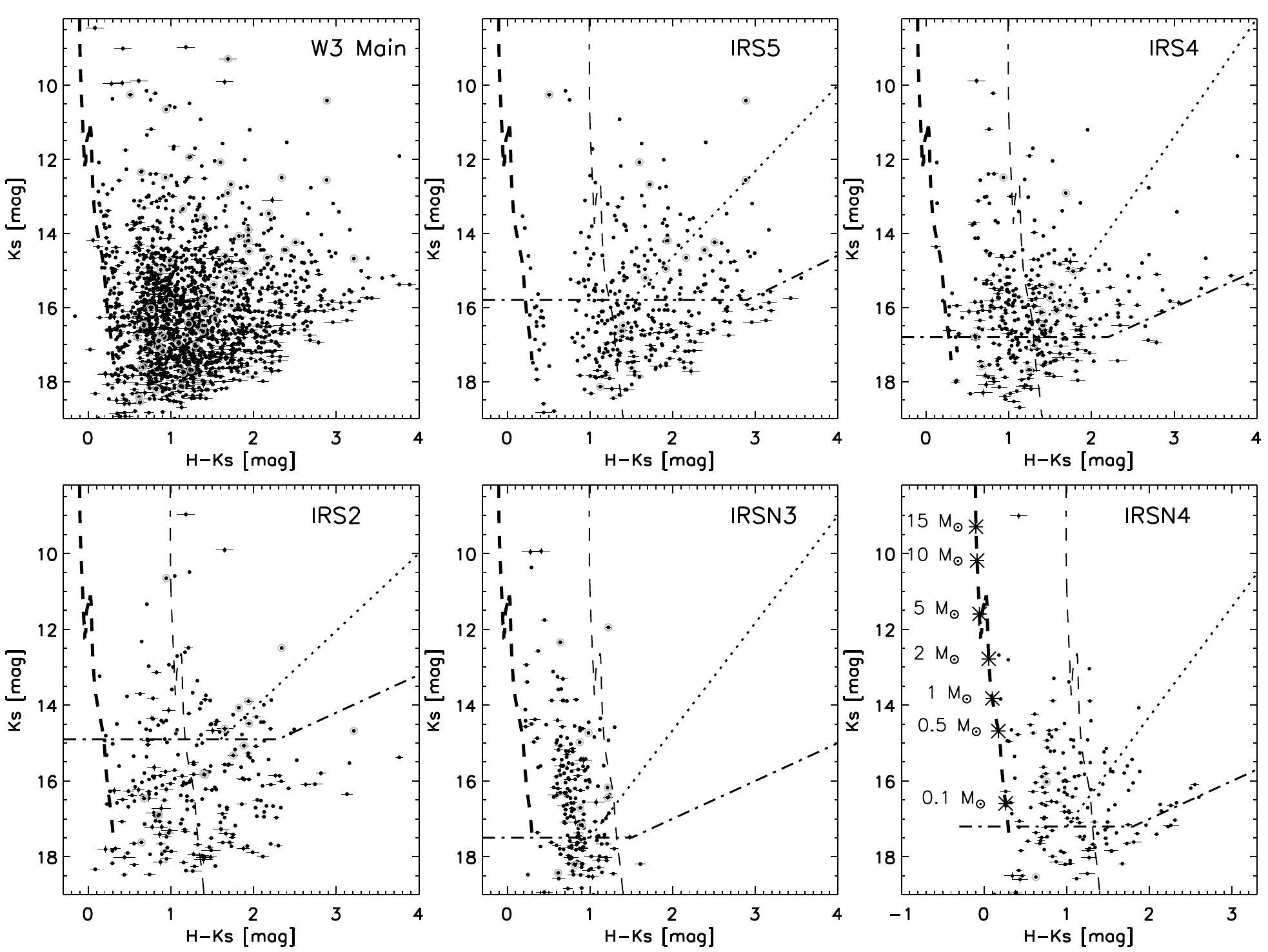}
   \caption{$H$-\Ks\ vs \Ks\ color-magnitude diagram of W3 Main and the chosen subregions in W3 Main. The subregions are circular areas covering the bright \HII\ regions with a radius of 1\arcmin, as shown in Fig \ref{fig:regions}. The objects marked with a gray circle are identified as \Ks-band excess sources using the method described in Sect. \ref{sec:excess}. No field subtraction has been performed in these plots. Note the different plotting range for the IRSN4 panel.
    Overplotted is the 2 Myr main sequence and PMS isochrone \citep{Brott11,Tognelli11} for \Ak = 0 mag (thick dashed line) and \Ak = 1.5 mag (thin dashed line). The IRSN4 panel shows the masses indicated on the isochrone. The dash-dotted lines are 50 \% completeness limits for the $H$- and \Ks-bands taken from Table \ref{tab:regions}. The dotted line represents an estimate of the completeness due to the matching with the $J$-band. Above this line, the $JH$\Ks\ matched sample occupies the CMD, below the line, only $H$\Ks\ matched sources are found.}\label{fig:cmd_HII}
       \end{figure*}

To study the stellar population of W3 Main in more detail, we subdivide the observed area of W3 Main in five subregions. The regions, with a radius of 1\arcmin, are selected such that they include the most prominent \HII\ regions in W3 Main (Fig. \ref{fig:regions}, Table \ref{tab:regions}). Two regions are centered on the ionizing stars of the diffuse \HII\ regions W3J (IRSN3)  and W3K (IRSN4). Two other regions cover the northern part of W3 Main: the region around IRS2 covers the compact \HII\ region W3 A, while the region around IRS4 covers the \HII\ regions W3C, W3D and W3 H. Finally, we choose the central area of W3 Main, which includes the deeply embedded cluster around IRS5 and several ultra compact \HII\ regions.  These subregions cover the entire evolutionary sequence of the \HII\ regions in W3 Main as described by \citet{Tieftrunk97}.  To study their  stellar populations  in detail, we constructed CMDs for all the regions (Fig. \ref{fig:cmd_HII}). In the CMDs, only the $H$\Ks\ matched catalogue is shown. The CMDs are plotted starting with the youngest region IRS5 and ending with the subregions covering the more evolved diffuse \HII\ regions around IRSN3 and IRSN4. 

The derived magnitudes at which the  50\% completeness is reached is varying strongly with location in the near-infrared images. To assess the photometric completeness in each subregion we calculate the brightest magnitude where a photometric completeness of 50\% is reached as derived from our artificial star experiments (Sect. 2.2). 

Using only stars above this magnitude limit we assure that in the entire area of the subregion the completeness is 50\% or better. Table \ref{tab:regions} summarizes the results showing that, in the IRS2 subregion, the completeness limit of 50\% is already reached at \Ks\ = 14.9 mag.  In Fig. \ref{fig:cmd_HII}, the dash-dotted lines represent the combination of the derived \Ks\ and the $H$-band completeness limits (Table \ref{tab:regions}).  The dotted lines show the effect of adding the $J$-band completeness limits and show that the detection of red sources in $JH$\Ks\ is very incomplete in our dataset.

The observed $H$-\Ks\ color spread  for the different subregions is due to  extinction variations.  Table \ref{tab:regionsprop} shows the average and the standard deviation of the  extinction towards the different subregions above \Ks\ = 14.9 mag after removing the foreground population with \Ak\ $<$ 0.4 mag.  We chose \Ks= 14.9 mag to be the limit at which the  entire stellar population is photometrically complete up to $H$ - \Ks\ = 3 mag. This covers the largest possible extinction range.  The youngest regions, IRS5 and IRS4, show the largest variations and the highest average extinction values. This is consistent with our derived extinction map (Fig. \ref{fig:akmap}) and the presence of cold, dense dust associated with these subregions \citep{Moore07,Wang12,RiveraIngraham13}. 
The IRSN3 subregion, associated with the more evolved \HII\ region W3 J, shows a very narrow observed color spread as well as average extinction, it demonstrates that most of the parental molecular cloud has been disrupted, consistent with the absence of molecular gas towards IRSN3 \citep{Tieftrunk98}.

\begin{table}
\caption{Summary of derived properties of subregions  in W3 Main.}             
\label{tab:regionsprop}      
\centering                   
\begin{tabular}{r c c c}       
\hline\hline                
Name  & \Ak & $\sigma$ \Ak & Slope KLF  \\ 
& mag & mag &  \\
\hline                        
W3 Main & 1.8 & 0.9 &  0.31 $\pm$ 0.05\\
IRS5		& 2.4 & 0.9 & 0.30 $\pm$ 0.13\\
IRS4		& 1.8 & 1.0 & 0.29 $\pm$ 0.10\\	
IRS2		& 1.8 & 0.8 & 0.28 $\pm$ 0.13\\
IRS N3	& 0.9 & 0.3 & 0.24 $\pm$ 0.15\\
IRS N4	& 1.2 & 0.5 & 0.14 $\pm$ 0.88\\ 
\hline                                  
\end{tabular}
\end{table}

\subsection{Disk fractions}\label{sec:excess}

\begin{figure}
\centering
\includegraphics[width=\columnwidth]{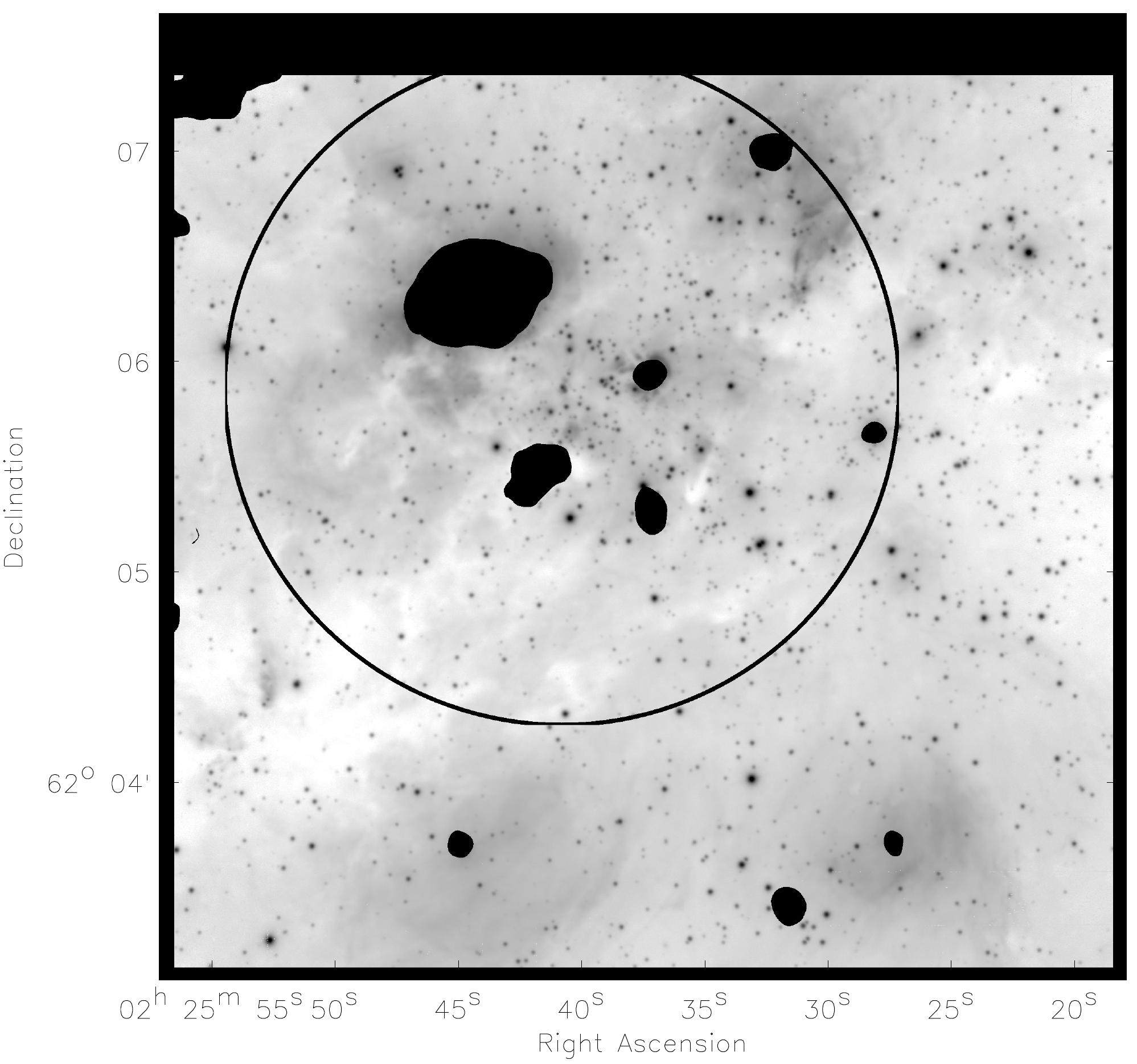}
\caption{Location of the two selected regions for the calculation of the disk fraction. The circle  marks the inner 1 pc centered on the position of IRS5. The masked out areas, plotted in black, have bright completeness limits and have been excluded from the disk fraction calculation enabling the calculation of the disk fraction for stars more massive than 0.51 \msun.\label{fig:disk_fractionmask}}
       \end{figure}

The presence of circumstellar disks  around young stars in W3 Main can be detected via an observable IR-excess. The near-infrared passbands are especially sensitive to the detection of the hot inner disks.  The fraction of sources surrounded by a circumstellar disk  strongly depends on the age  of the system as well as the amount of UV radiation present in the cluster. \citep{Hernandez08}. Searching for variations in the disk fraction in W3 Main would allow us to study the effect of the age spread and the massive stars on the disk fraction.  To look for radial variations in the disk fraction, we split the observed area of W3 Main in two regions. We calculate the disk fraction for the central 1 pc  around IRS5 as well as for the region outside this central parsec (Fig. \ref{fig:disk_fractionmask}).

\begin{figure}
\centering
\includegraphics[width=\columnwidth]{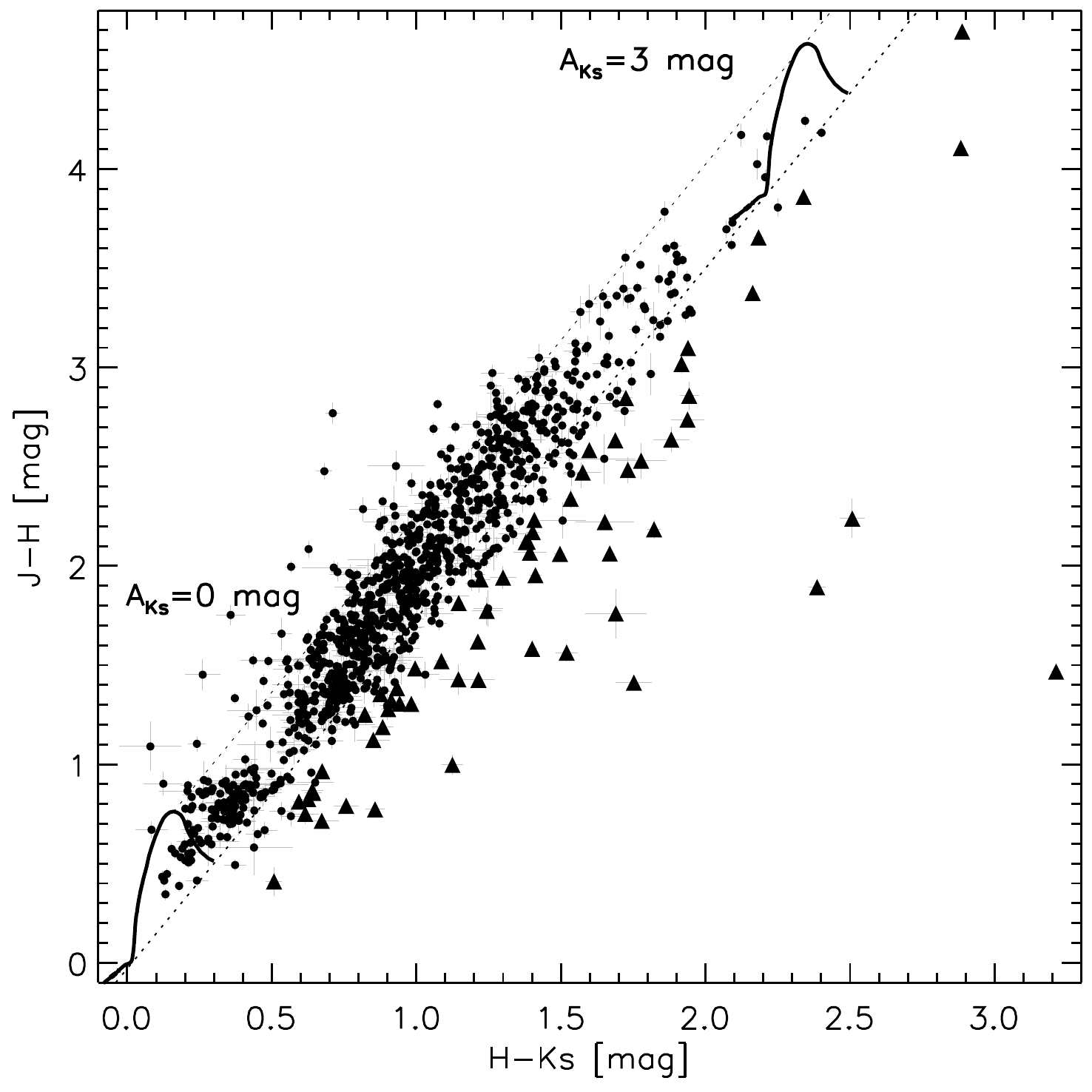}
\caption{ $H$-\Ks\ vs $J$-$H$  color-color diagram of W3 Main. Overplotted are the theoretical 2 Myr isochrones for \Ak=0 and \Ak=3 mag (solid lines). The diagonal dotted lines represent the expected location of reddened stars without excess (filled circles). The stars selected for possessing a near-infrared excess (triangles) are selected  to be to the right of this reddened main sequence with more than 3 times their color-error in both $J$-$H$ and $H$-\Ks.}\label{fig:CCD}
       \end{figure}

Using the $J$-$H$ vs. $H$-\Ks\ color-color diagram (CCD, Fig. \ref{fig:CCD}), we selected the sources which are 3 times their color-error  in both $J$-$H$ and $H$-\Ks\  away from the reddening vector of main sequence stars as infrared excess sources. Based on this analysis in the CCD, a total of 63 IR excess sources are found. However, to derive the excess fraction we have to correct  for photometric completeness.

Due to the presence of the emission of the inner-disk, IR-excess sources will appear brighter than the non-excess sources with the same mass.  To take this into account, we assume that the shortest wavelength ($J$-band) is the least affected by the circumstellar disk and the measured $J$-band flux is originating from the star.

To calculate the disk fraction  we apply the following steps: First, only the stars with an associated completeness above 50\% as derived in Sect. \ref{sec:completeness} are selected. Applying the strict  completeness limits from Table \ref{tab:regions} guarantees a complete sample of the entire observed field, but results in a rather high minimum mass ($M$ = 1.4 \msun) for the disk fraction calculation. By removing small areas, where the completeness limits are very bright,  the 50 \% completeness limits can be much fainter, allowing us to probe lower masses. For the disk fraction analysis we lower the completeness limits to $J$=20.1 mag, $H$= 18.9 mag and \Ks\ = 16.3 mag.  This results in the exclusion of some small regions centered on the HII regions around IRS2 and IRS3a and several bright stars (Fig. \ref{fig:disk_fractionmask}), but yields  a disk fraction complete down to $M$ = 0.51 \msun.

Second, we remove the fore- and background stars. The field contamination is estimated using the control field observations. The foreground stars were selected by having \Ak\ $< 0.4$ mag. The background stars behind the cluster will typically have a higher extinction than the background stars observed in the control field as the extinction inside the cluster is not present in the control field. 

 The CCD of the control field (Fig. \ref{fig:cmd_control}) shows a similar pattern as the science field. A blue foreground sequence is visible ($H$-\Ks $<$ 0.4) with a reddened sequence along the reddening line. The sources bluewards of $H$-\Ks $<$ 0.4 we consider foreground sources along the line of sight towards W3 Main. The sources redwards are most likely in the background.  To correct for the reddening of background sources in the control field, we de-reddened all the sources redwards of  $H$-\Ks $=$ 0.4 mag back to $H$-\Ks $=$ 0.4 mag, assuming that $H$-\Ks $=$ 0.4 is the reddening caused by the material in front of W3 Main.

To estimate the contribution of the expected background stars in the region of the cluster, we artificially redden the de-reddened control field stars (with \Ak $> 0.4$ mag) with a random extinction drawn from the extinction distribution  after removing the 0.4 mag foreground extinction  towards each subregion. This results in a more realistic reddening of the background stars in the region of the cluster. After this, both the foreground and  background star contamination are removed on a statistical basis using a Monte-Carlo method  \citep{Schmalzl08} in color-magnitude space.  

 As noted in Sect. \ref{sec:extinction} in some of the most deeply embedded areas around IRS5 and IRS4 the extinction is so high that most likely not even all the cluster members are detected. Here the background contamination would not be detectable either. To investigate whether the background subtraction strongly affects the disk fraction, we additionally calculate the disk fraction with only removing the foreground contamination with \Ak\ $< 0.4$ mag.

 The extinction towards the excess sources is derived from the extinction map (Sect. \ref{sec:extinction}).
We calculate the extinction map from the full $JH$\Ks\ catalogue with the excess sources removed in the same manner as Sect. \ref{sec:extinction}. The resulting extinction on the position of the excess source is applied to the source.

\begin{table*}
\caption{Disk fractions in W3 Main}             
\label{tab:diskfractionprop}      
\centering                   
\begin{tabular}{r | c c c | c c c}       
& \multicolumn{3}{c}{No background subtraction} & \multicolumn{3}{c}{Background subtracted}\\
\hline\hline        
region  & N$_{disk}$\tablefootmark{a} & N$_{diskless}$\tablefootmark{a} & disk fraction (\%) &  N$_{disk}$\tablefootmark{a}& N$_{diskless}$\tablefootmark{a} & disk fraction (\%) \\
\hline                        
W3 Main  & 19.3 & 282.1 &  6.8 $\pm$ 1.6 & 18.3& 236.6 & 7.7 $\pm$ 2.3\\
r $<$ 1pc & 12.1 & 129.5 & 9.4 $\pm$ 2.8  & 12.2& 125.4 & 9.4 $\pm$ 3.0 \\
r $>$ 1pc & 7.1 & 156.7 &  4.5 $\pm$ 1.7  & 7.1 &  128.3 &  5.6 $\pm$ 2.2\\
\hline                                  
\end{tabular}
\tablefoot{
\tablefoottext{a} Corrected for photometric incompleteness.
}
\end{table*}



With the new, slightly less conservative completeness limits  we are complete for stars more massive than M = 0.51 \msun\ with an \Ak\ $<$ 1.5 mag ($J$ = 20.1 mag at \Ak\ = 1.5 mag). After applying this selection and applying the $J$-band completeness correction for each star, the disk fractions are calculated, as this is the most limiting completeness factor. From the 63 detected excess sources, 19 source remain after applying all the selection criteria. The other sources are either too faint and fall below the completeness limits or have an extinction higher than 1.5 mag. In Table \ref{tab:diskfractionprop} the number of excess and non-excess sources are listed after they are completeness corrected. 

The resulting disk fractions are presented in Table \ref{tab:diskfractionprop}. The results are presented with and without background source subtraction.  The total disk fraction of W3 Main is 6.8 $\pm$ 1.6 \% when taking the background subtracted sample and 7.7 $\pm$ 2.3 \% when the background stars are not subtracted. This shows that the effect of the background stars is minimal and does not change the disk fraction significantly. Additionally, we find a suggestive change in disk fraction from the inner 1 pc (9.4 $\pm$ 3.0 \%) to the outer area (4.5 $\pm$ 1.7 \% or 5.6 $\pm$ 2.2). The difference between these two disk fractions is less than 2$\sigma$, but would be consistent with the suspected younger age from the presence of the hyper-compact and ultra-compact \HII\ regions in the central parsec.

The more recent disk fraction studies are mostly performed using $L$-band data to identify the excess fraction, as it has been proven to be more reliable \citep{Haisch00} by detecting more excess sources than using the \Ks-band as indicator.
 However, in the case of W3 Main, the Spitzer/IRAC data does not reveal many point sources as the bright diffuse emission of the \HII\ regions and the low-spatial resolution hamper the detection of faint point sources \citep{Ruch07}. Therefore, our analysis is limited to the $JH$\Ks\ data of W3 Main.  Comparing the disk fractions derived from \Ks-band excess with the  $L$-band based studies is challenging as $K$-band disk fractions are typically significantly lower \citep[e.g.][]{Haisch00}. Using literature data, however, \citet{Yasui09,Yasui10} determined the disk fraction derived from $K$-band excess of several young stellar clusters covering an age range from 0 - 10 Myr.  They show that the disk fractions derived from $K$-band data are about a factor of 0.6 lower than the fraction derived from $L$-band data and slightly more scattered on the disk-fraction vs age diagram.  Applying this conversion factor, the total disk fraction of W3 Main would be between 11.3 and 12.8 \%\ (with and without background subtraction), with the central 1 pc raising to 15.7 $\pm$ 5.0 \%.

With an estimated age of $\sim$3 Myr or less, the measured disk fraction of W3 Main is still rather low compared to other embedded clusters of the same age \citep[30 - 50 \%,][]{Hernandez08}.  Similarly low disk fractions, however, are observed for clusters possessing stars of spectral type O5V and earlier. \citet{Fang12} show that in the case of clusters containing these types of massive stars the disk fraction is significantly faster declining with age. They find a disk-fraction dependance of age as $f_{\rm{disk}} = e^{-t/1.0}$, with $t$ being the age in Myr, for the clusters harboring very massive stars. This is significantly faster than for cluster without these very massive stars ( $f_{\rm{disk}} = e^{-t/2.3}$). \citet{Fang12} derived these relations from disk fractions calculated for stars more massive than 0.5 \msun. For W3 Main we have calculated the disk fraction with the same mass limit, allowing a fair comparison with these results.  The converted $L$-band disk fraction of 11.3 - 12.8 \% fits very well with the fast declining disk fraction found for clusters containing very massive stars. The most massive star in W3 main, IRS2, is currently classified as a O6.5V - O7.5V, however, as it is already evolved from the zero-age-main sequence, the current mass derived from its position in the Herzsprung Russell diagram (35 \msun) suggests its spectral type was around O5V when it was on zero-age-main sequence \citep{Bik12}. \citet{Fang12} attribute the decline in disk fraction to the FUV radiation of the massive stars  which heats up the disk of the low-mass stars and enhances the accretion onto the star via increased magneto-rotational-induced turbulence \citep{Stone00}, making the disks evolve faster and therefore short-lived.


\subsection{Luminosity functions}\label{sec:lumfunction}

   \begin{figure*}
   \centering
  \includegraphics[width=16cm]{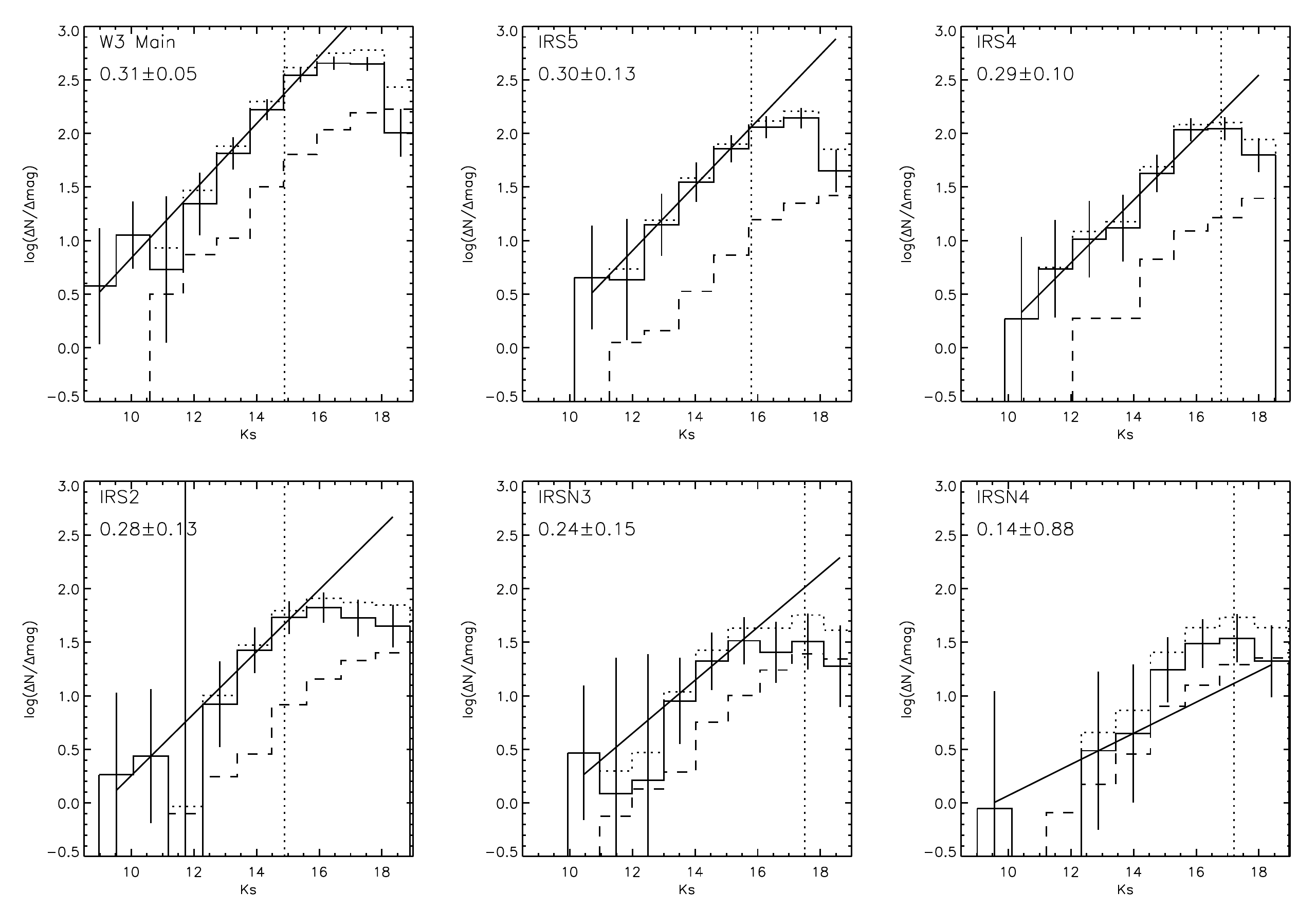}
   \caption{\Ks-band luminosity functions for the entire W3 Main region and the different subregions  constructed from the \Ks-detected sources only. The dotted histograms are the observed sources binned with a bin size of 1 magnitude. The dashed histograms represent the histogram of the  stars in the control field, scaled to the area of the corresponding subregion. The  field-corrected histogram is plotted with a solid line. The vertical dotted lines represent the magnitude above which we are photometrically complete with a completeness of at least 50\%. A linear fit yields the slopes indicated in the figures. }\label{fig:KLF}
       \end{figure*}

Using the \Ks-only catalogue we constructed \Ks-band luminosity functions (KLFs) of the five subregions  of W3 Main as defined in Sect \ref{sec:HIIregions}.  The \Ks\ only sample is the deepest sample and probes the highest extinction as it is not limited by extinction losses in the shorter wavelength detections. This implies that no correction for extinction can be applied as no color information is available for the reddest and faintest sources. However, the shape of the KLF  holds valuable information on the underlying IMF, cluster age and star formation history \citep{Muench00}. 

Fig. \ref{fig:KLF} shows the KLFs for W3 Main as a whole and the five subregions with a 1 magnitude binning. 
The KLF is constructed by correcting for the photometric completeness for each star as derived from the 2 dimensional \Ks\ completeness map. We only used the stars which have a photometric completeness above 50\% (Table \ref{tab:regions}).  To correct for the field contamination we applied the method described in Sect. 3.4.  As the completeness limits in the control field are a lot deeper, the matched $H$\Ks\ catalogue of the control field still provides a good estimate of the control field of the \Ks\ only catalogue of W3 Main.
We artificially redden the  extinction-corrected background stars in the control field with the reddening distribution of the corresponding sub region and construct the luminosity function of the reddened control field. After scaling to the same observed area (Table \ref{tab:regions}), the KLF of the control field is subtracted  from the observed KLF of W3 Main. By comparing the observed KLF with that of the field subtracted KLF, it becomes clear that the contamination by the field is small and does not affect the measured slopes of the KLFs.

The flattening of the KLFs at magnitudes fainter than the completeness limits is caused by the incompleteness of the data.  However, the flattening of the KLF of IRSN3 happens at brighter magnitudes (\Ks\ = 15 mag) than the incompleteness limit and is caused by the change of slope of the underlying mass function \citet{Muench00}. The position of the break   is consistent with the model KLFs of \citet{Muench00} with an age of 1-3 Myr.
We fitted a power law to the  field-corrected KLFs above \Ks\ = 14.9 mag, the brightest completeness limit of all subregions, to determine the slope  of the luminosity function. All KLFs are fitted to the same limiting magnitude, allowing a fair comparison of the slopes. 

The fits were performed on the KLFs binned with a 1 magnitude binning. Poissonian statistics were considered as uncertainties in number of sources per magnitude bin. Considering that there are biases introduced in uniformly binned data  \citep{Maiz05}, we also counted the stars in variable-sized bins, resulting to equal stellar numbers per magnitude bin. The derived KLF slopes are found to be identical in both fitting processes. The slopes, and 1-sigma errors, of the KLFs are given in Table \ref{tab:regionsprop} and are all consistent with a slope of 0.3.  The slope of IRSN4 is very poorly determined due to the lack of bright sources. Furthermore, the slopes are similar to the results found by  \citet{Ojha04}  towards W3 Main.

\subsection{Mass function}\label{sec:massfunction}

\begin{figure*}
\includegraphics[width=18.3cm]{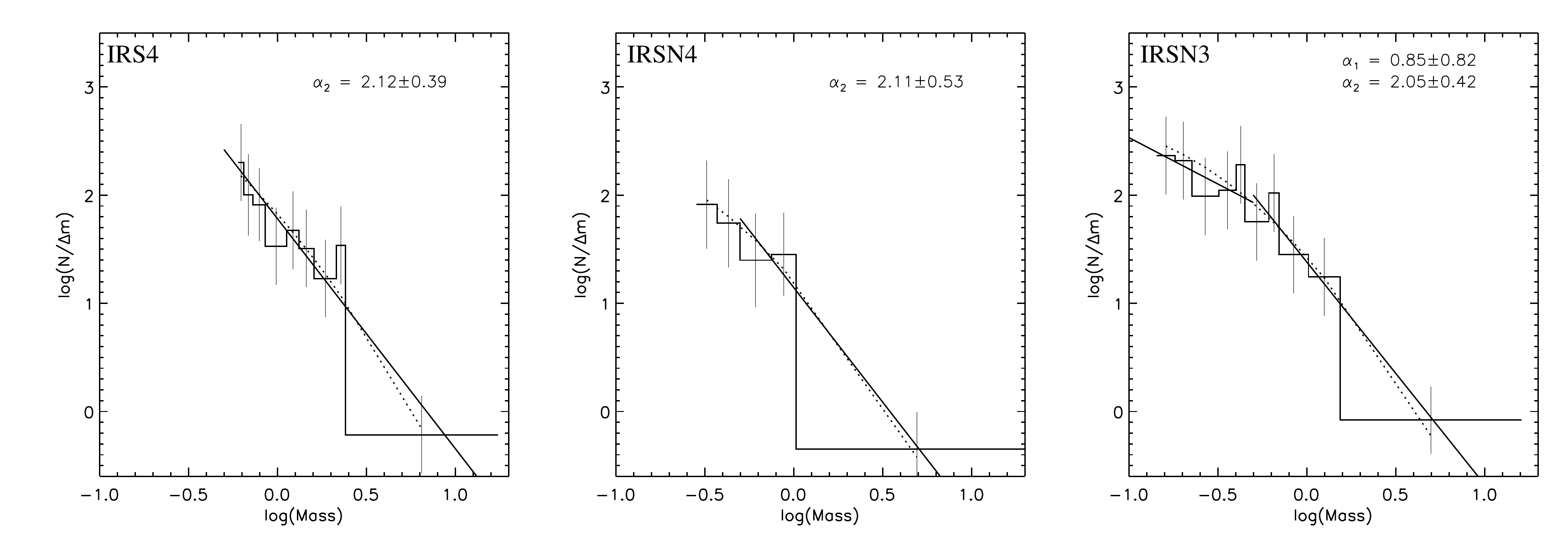}
\caption{Mass functions of the subregions  IRS4, IRSN3 and IRSN4. Power laws are fitted to the mass functions above 0.5 \msun\ (solid lines) and fore IRSN3 an additional power-law is fitted to the lower-mass end (0.06 \msun\ - 0.5 \msun). Overplotted  is also the  log-normal system mass function representation of \citet[dotted line]{Chabrier05}.}\label{fig:massfunctions}
\end{figure*}

In clusters where the extinction is  homogeneous, the KLF can be converted into a luminosity function using the mass-to-light-ratio. However, in W3 Main the variable extinction prevents this as no unique relation between the observed \Ks\ magnitude and mass exists. Instead, we use the color-magnitude diagram with the $H$\Ks\ sample to construct the mass functions of the different subregions.
First  the fore- and background star contamination are removed from our $H$\Ks\ catalogue on a statistical basis using the Monte-Carlo method described in Sect. \ref{sec:excess} and by taking into account the extra extinction that is added to the background stars to represent the cluster line of sight.
The mass of the stars is calculated by de-reddening the sources using the \citet{Nishiyama09} extinction law.  The reddening corrected magnitudes are compared to the  isochrones calculated by  \citet{Brott11} for the stars above 5 \msun. The PMS tracks of \citet{Tognelli11}  are used for the determination of the mass between 0.06 and 5 \msun. 

A comparison between the location of the PMS stars in the CMD with theoretical isochrones is needed to estimate their masses. However, the position of the PMS stars can be altered by different accretion histories, variability or unresolved binarity \citep[e.g.][]{Preibisch12}. This will have an effect on the resulting slope of the mass function.  Additionally, the mass assigned to the stars in the cluster will depend on the chosen age of the isochrones. As an age spread is present in W3 Main, we calculate the mass function with the 1, 2 and 3 Myr isochrones to examine whether the slope of the mass function varies significantly with the chosen isochrone age. Additionally, we calculate the mass functions by randomly assigning an age to each star choosing from the 1, 2 and 3 Myr isochrones. 

 In the transition region, where the PMS merges into  the main sequence, no unambiguous mass determination is possible. Here we chose the mass of the PMS isochrone instead of the main sequence isochrone. For the mass function this has no effect as this mass range is part of the highest mass bin, which covers a large range of masses as only a few objects are present. 
 
The range in mass and extinction over which the mass functions can be constructed depends on the photometric completeness in the $H$ and \Ks\ bands. At high values of \Ak, the mass limit where we are complete is very high, therefore we limit the construction of the mass function to relatively low \Ak (Table \ref{tab:massfunctions}). Using the 50\% completeness limits shown in Table \ref{tab:regions}, the limiting masses are 0.14 \msun\ for IRSN3, 0.28 \msun\ for IRSN4 and 0.59 \msun\ for the IRS4 subregion (Table\ref{tab:massfunctions}).  For the subregions IRS2 and IRS5 we can not create a statistically meaningful mass function, as due to the very bright 50\% completeness limit, the mass range which could be covered in these  subregions  was not enough to create a meaningful mass function.

 Following \citet{Maiz05} we construct the mass functions for the subregions IRS4, IRSN3 and IRSN4 using a variable bin size ensuring an equal number of stars per bin (10 stars per bin). While constructing the mass function, each star is corrected for its \Ks\ incompleteness as derived from the 2D incompleteness analysis. The \Ks\ incompleteness is the dominant incompleteness factor. The resulting mass functions, plotted in linear units, are shown in Fig. \ref{fig:massfunctions}.

\begin{table}
\caption{Parameters of the mass functions derived in W3 Main.}             
\label{tab:massfunctions}      
\centering                   
\begin{tabular}{r c c c c}       
\hline\hline                
Name  & Min. Mass & Max. Mass & max \Ak\ & Slope \\ 
	& (\msun)& (\msun) & (mag) & ($>$ 0.5 \msun)\\
\hline                        
IRS4		& 0.59 & 17.3 & 2.3& 2.12 $\pm$ 0.39 \\	
IRS N4	& 0.28 & 23.5 &1.8&  2.11 $\pm$ 0.53  \\ 
IRS N3	& 0.14 & 16.0 & 1.3& 2.05 $\pm$ 0.42\\
\hline                                  
\end{tabular}
\end{table}

For each of the regions we fit the mass function slope for masses above 0.5 \msun\ and found values of  $\alpha \sim 2.1$ (Table \ref{tab:massfunctions}), similar to a \citet{Salpeter55} slope of $\alpha$ = 2.3.  For the IRSN3 mass function we also fitted the mass range below 0.5 \msun\ and found that the lower-mass end has a flatter slope ($\alpha = 0.85 \pm 0.82$).  The derived values are very similar to those of  the \citet{Kroupa01}  mass function, showing a broken power law with a slope $\alpha_{1}$ = 1.3 between 0.08 and 0.5 \msun\ and $\alpha_{2}$  = 2.3  above 0.5 \msun\ (log \msun\ = -0.3).  Our derived mass function is also consistent with the  log-normal system mass function of  \citet{Chabrier05}.

The absolute error on the mass is hard to quantify as it depends on the chosen extinction law, distance uncertainty and age of the cluster.
The errors invoked by the extinction law and distance result in a shift to higher or lower-masses of the entire population. The error due to the age of the cluster however changes the slope of the mass function. For the mass functions described above, the age of each star was chosen randomly to be 1, 2 or 3 Myr, representing the age spread found in W3 Main. Due to the uncertainty in the individual ages for each region, we consider the mixed age slopes the best representation of the mass functions of this region. To quantify the effect of the cluster age on the mass function we also calculate the slopes  choosing a fixed age for the stellar population of subregion IRSN3.

The slopes of the mass functions with the 2 and 3 Myr isochrones do not significantly differ from the mass function constructed by randomly assigning an age to each star. The mass function made with the 1 Myr isochrone, however, becomes more similar to  a single power law from 0.14 to 15 \msun, the upper end is flatter ($\alpha_{2}$ = 1.75 $\pm$ 0.52), while the slope of the mass function below 0.5 \msun\ becomes steeper ($\alpha_{1}$ = 1.56 $\pm$ 0.52). This can be explained by a change in  slope of the low-mass end of the mass-luminosity relation \citep[e.g.][]{Zinnecker93}. 

\section{Discussion}\label{sec:discussion}

In this paper we  perform an analysis of the  $JH$\Ks\ imaging data of W3 Main. After a detailed completeness analysis we  statistically determine several properties of the stellar population of W3 Main. We derive a two dimensional extinction map towards W3 Main, showing the presence of two high-extinction regions towards the center of W3 Main. We derive the disk fraction, K-band  luminosity function and mass functions for several subregions in W3 Main.  In this section we use these properties to derive constraints on  age spread in this cluster, possible dynamical evolution as well as on possible formation scenarios of W3 Main.


\subsection{Age spread}

Analyzing the massive stars and the \HII\ regions around them, evidence is presented that the massive stars in W3 Main have been forming over the last 2-3 Myr \citep{Tieftrunk97,Feigelson08,Bik12}.  \citet{Feigelson08} inferred from the very low fraction of Spitzer IR excess among X-ray selected stars that most of the W3 Main cluster stars distributed over several pc region are several million years old. The age estimate of the most massive star IRS2 (2-3 Myr) and the presence of the very young hyper- and ultra-compact \HII\ regions suggests that star formation in W3 Main is still ongoing. Additionally, the detailed analysis of the mm cores detected with SCUBA using SMA observations shows that two of them are actively forming stars and driving outflows. The youngest of the cores shows no star formation activity yet and is on the onset of star formation \citep{Wang12}.

With our analysis of the properties of the low-mass stars we can determine whether there is any evidence for differences in evolutionary phase.  The extinction maps and the (sub)mm maps \citep{Moore07,RiveraIngraham13} show the presence of two dense molecular cores, centered on IRS5 and IRS4 (Fig. \ref{fig:akmap}). The general stellar population in the regions around these two massive stars shows a high average extinction (\Ak=2.3 mag  and 1.7 mag respectively) as well as a large spread in measured extinction values (Table \ref{tab:regionsprop}). The subregions around IRSN3 and IRSN4 show significantly lower extinction as well as smaller extinction spread. 

 The difference in disk fraction between the central parsec around IRS5 (9.4 $\pm$ 3.0) and the surrounding area (4.5 $\pm$ 1.7 or 5.6 $\pm$ 2.2, without or with field subtraction) is suggestive for an evolutionarily difference between the two regions. This would be consistent with the detected age spread. The shapes of the luminosity functions support this finding as well. The very shallow slope of the KLF towards the cluster around IRS5, observed by \citet{Megeath96} and \citet{Ojha04}, was interpreted as evidence for a very young stellar population ($\sim$0.3 Myr). The location of the break in the KLF of the subregion IRSN3 suggests an age of a few Myr, indicating an older age than the central area of W3 Main.

Based on this evidence we conclude that the observed age spread in the massive stars might be applicable to the low-mass stars as well.  \citet{Feigelson08} concluded, based on their X-ray observations and the lack of Spitzer excess sources, that the PMS stars might be significantly older than the massive stars. However, the above results would suggest that the age spread detected for the high-mass stars is also present in the low-mass stellar population.
 
Evidence for age spreads is found in other young stellar clusters. While \citet{daRioOrion10} and \citet{Reggiani11} deduced and age spread of several Myr for ONC, upper limits on the age spread in the more massive and compact starburst clusters in NGC 3603 and Westerlund 1 are 0.1 Myr and 0.4 Myr, respectively \citep{Kudryavtseva12}. This could suggest differences in the formation mechanism between less massive star clusters like W3 Main and the massive, very compact starburst clusters.


\subsection{Cluster dynamics}

The spherical nature of W3 Main as seen in the X-ray data \citep{Feigelson08} suggests that the cluster is already  dynamically evolved while still forming stars. The density enhancement towards IRS4 indicates the presence of some sub-clustering in W3 Main. As IRS4 is one of the younger regions, the recently formed stellar population might be responsible for this sub-clustering.  Another  sign of dynamical evolution is dynamical mass segregation where the massive stars preferentially move to the center of the cluster and the low-mass stars migrate to the outskirts \citep{Allison09, Harfst10}. Simulations show that this process can be effective for the more massive stars in a few Myr \citep{Allison09}. This dynamical mass segregation would lead to radial variations in the  slope of the mass function  as observed in other clusters \citep[e.g.][]{Gouliermis04,Brandner08,Harayama08,Gennaro11,Habibi13}.

We  derived the mass functions for a mass and extinction limited sample of the stellar population of the different subregions in W3 Main. This ensures that the mass function samples the complete stellar population in this mass and extinction range. All mass functions derived for the different subregions have  a similar slope  within the uncertainties of 0.3. The high-mass end of all mass functions have slopes consistent with a \citet{Salpeter55} slope. Additionally,  in the subregion around IRSN3 where we could probe the stellar content to 0.14 \msun, we found evidence for the break of the mass function around 0.5 \msun, as seen in the \citet{Kroupa01}  and \citet{Chabrier05} representations of the IMF.

In the range of masses and extinctions in which we can probe the mass function, we do not see any spatial variations in the slope of the IMF. Also the shape of the mass function of IRSN3, in which very low-mass stars are detected, does not display any evidence for mass segregation. \citet{Ojha09}, however, presented evidence that the very low-mass stars are segregated. They detect a lack of the lowest mass stars (M $<$ 0.1 \msun) towards the central region of the cluster.   Due to the extremely high extinction towards this  area we do not have a complete sampling of these low-mass stars in our data. \citet{Ojha09} used a constant extinction correction towards the entire cluster (\Ak = 1.5 mag). This underestimates the extinction toward the central area around IRS5 and might  be the reason why \citet{Ojha09} did not detect stars below 0.1 \msun.   The extinction toward this area  (\Ak $>$ 2 mag) as measured in this work is
much higher than the average extinction towards W3 Main adopted by \citet{Ojha09}.

\subsection{Cluster formation}

With all the information acquired on the stellar content  discussed in this paper as well as the X-ray observations \citep{Feigelson08}, and the deeply embedded star-forming cores \citep{RiveraIngraham13}, we can now construct a possible scenario of how clusters like W3 Main form. 

W3 Main is a complex region, part of an even larger conglomerate of star-forming regions in the W3/W4/W5 complex \citep{Megeath08}. Its location suggests that W3 Main is triggered by the expanding bubble around the OB association IC 1795  \citep{OeyW305}. However, based on the spherical distribution of PMS stars detected in the X-ray data, \citet{Feigelson08} concluded that external triggering by the OB association might be less likely.  Nevertheless, according to these authors, the different generations of massive stars present in W3 Main \citep{Tieftrunk97,Bik12} suggest that internal triggering could have played a role in W3 Main. By studying the molecular CO gas distribution around the \HII\ regions, \citet{Wang12} found evidence that IRS5 is currently forming in a  layer of molecular gas swept-up by W3 A, the \HII\ region around the most massive star IRS2. 

In this study of the low-mass stellar content we find evidence that not only the massive stars show an age spread, but also  the low-mass stellar content shows evidence for such an age spread. The youngest regions appear to be located towards the center of W3 Main. Deeply embedded molecular cores are found here, analyzed in detail using Herschel observations \citep{RiveraIngraham13}. 

Simulations indicate that the formation process and the early life time of an embedded cluster might be a very dynamical process \citep[e.g.][]{Dale11,Pelupessy12}. Recently formed stars interact with molecular cores still present in the molecular cloud in which the cluster is forming. These simulations also show that the ionizing feedback from massive stars is of limited effect. \citet{Smith09} and \citet{Dale11} found that the dense molecular cores are unaffected by the ionizing radiation of the massive stars.  From our observations in W3 Main we find  that the ionizing feedback  of  the massive stars on the  molecular cores is only effective over a relatively small distance \citep{Peters10}. The most luminous \HII\ region, harboring the most massive star IRS2,  has a size of  $\sim$ 0.5 pc only, suggesting that in the past 2-3 Myr the ionizing feedback of IRS2 has only been effective within these 0.5 pc. Also the higher disk fraction, as compared to the outer regions, towards the IRS5 subregion where many O stars are forming demonstrates the limited influence of destructive feedback in W3 Main. At spatial scales beyond 0.5 pc dense molecular cores, still in the formation process, can co-exist with OB stars without being destroyed.  The low disk fraction could suggest that the effect of the FUV radiation is penetrating deeper into the cluster heating up the molecular material and the disks around the low-mass stars \citep{Fang12}.

The age of the most massive O star IRS2, and the nature of the \HII\ regions suggest that  star formation in W3 Main started at least 2-3 Myr ago, and due to dynamical  interactions  with dense molecular cores in the surroundings and possible internal triggering, star formation is continuing until the present day. Currently the youngest population is located in the center and may be the latest dense concentration of molecular gas in the contracting cloud, forming the youngest sub cluster around IRS5. 

What we observe of W3 Main is a two dimensional projection of the three dimensional structure of the cluster. As observed towards the Orion nebula, the youngest region (BN object) is located behind the Orion Nebula Cluster, demonstrating that we need to be careful in interpreting the relative location of  OB stars and molecular cores in embedded clusters.  Proper motions to be obtained with e.g. GAIA  and radial velocities in combination with detailed simulations of the stars and  gas simultaneously \citep[e.g.][]{Pelupessy12} are additionally needed to make quantitative statements on the dynamical evolution of embedded clusters like W3 Main.

\section{Summary and conclusions}\label{sec:conclusions}

Using deep LUCI1/ LBT near-infrared photometry we have performed a detailed study of the entire stellar content of the young embedded cluster W3 Main. A two dimensional completeness analysis using artificial star experiments is a crucial tool to properly assess the spatially variable completeness limits. We derive the infrared excess fraction, mass functions and \Ks-band luminosity functions for several subregions in W3 Main, allowing the search for evolutionary differences among them as already found for the massive stars by  \citet{Bik12}.  Our conclusions can be summarized as follows.

\begin{itemize}
\item{We create a 2-dimensional extinction map based on the observed colors of the low-mass stellar content of W3 Main. We find the extinction to be highly variable on small spatial scales. The highest extinction areas are found in the center of W3 Main towards the high-mass protostars IRS5 and IRS4, consistent with (sub) mm observations.}
\item{We find that the apparent sub-structure visible in the near-infrared images is caused by the extinction variations. The stellar distribution as observed in the X-rays, where higher extinctions are probed, shows the real, almost spherical spatial distribution. The extinction alters that distribution when looking at the stellar population in the \Ks-band. While we attribute most of the sub-clustering observed at NIR wavelengths
   to these extinction effects, a true density enhancement appears to be detected
   towards the young region IRS4.}
\item{A  low \Ks-band excess fraction is found for W3 Main (6.8$\pm$ 1.6 to 7.7 $\pm$ 2.3). The low overall disk fraction is consistent with a fast disk evolution due to the FUV radiation of the most massive stars in W3 Main. We find a likely radial decline in the disk fraction from 9.4$\pm$ 3.0 \% in the central parsec of W3 Main to 4.5 $\pm$ 1.7 or 5.6 $\pm$ 2.2 in the outer regions. This would be consistent with the suspected younger age of the central area of W3 Main.}
\item{The \Ks-band luminosity functions  of the different subregions in W3 Main show a power law with a slope of $\sim$ 0.3 for all the subregions. We identify a break in the KLF of the IRSN3 subregion, consistent with the break in the mass function at $\sim$0.5 \msun.}
\item{We derived the mass functions for 3 subregions. The slope of the mass functions are consistent with a Salpeter slope. For the subregion around IRSN3 we construct the mass function down to 0.14 \msun. This mass function is consistent with a broken power law as proposed by \citet{Kroupa01} with the break at $\sim$0.5 \msun\ and with the \citet{Chabrier05} IMF. We find no evidence for mass segregation in contrast to previous publications.}
\item{The age of the most massive O star IRS2, and the nature of the \HII\ regions suggest that the star formation in W3 Main started at least 2-3 Myr ago, and due to dynamical interactions with dense molecular cores in the surroundings and  internal triggering by the expanding \HII\ regions, star formation is continuing until the present day.}
\end{itemize}

\begin{acknowledgements}
We would like to thank the anonymous referee for  careful reading and helpful suggestions which have improved the paper significantly. We thank Nancy Ageorges and Walter Seifert and the LBTO staff for their support during the observations. We thank Leisa Townsley and Eric Feigelson for sharing their X-ray catalogue of W3 Main and discussions. A.S. acknowledges generous funding from the German science foundation (DFG) Emmy Noether program under grant STO 496-3/1. D.A.G. kindly acknowledges financial support by the German 
Research Foundation (DFG) through grant GO\,1659/3-1. 

\end{acknowledgements}

\bibliographystyle{aa}
\bibliography{arjanbibLUCIW3pms}

\end{document}